\documentclass[pra,11pt,letterpaper,nofootinbib,showpacs,showkeys,eqsecnum
] {revtex4}
\usepackage{amsmath,amssymb,amsfonts,graphicx,hyperref}

\def\ket#1{ | #1 \rangle}
\def\bra#1{{\langle #1 | }}
\def\tr{ {\rm{Tr }}}
\def\vec#1{\boldsymbol{#1}}

\newtheorem{theorem}{Theorem}

\begin{document}

\title{Constrained bounds on measures of entanglement }

\author{Animesh Datta}
\email{animesh@unm.edu}
\author{Steven T. Flammia}
\author{Anil Shaji}
\author{Carlton M. Caves}
\affiliation{Department of Physics and Astronomy, University of
New Mexico, Albuquerque, New Mexico 87131-1156, USA.}

\date{May 31, 2007}

\begin{abstract}
Entanglement measures constructed from two positive, but not
completely positive maps on density operators are used as constraints
in placing bounds on the entanglement of formation, the tangle, and
the concurrence of $4 \times N$ mixed states. The maps are the
partial transpose map and the $\Phi$-map introduced by Breuer [H.-P.
Breuer, Phys.\ Rev.\ Lett.\ \textbf{97}, 080501 (2006)]. The
norm-based entanglement measures constructed from these two maps,
called negativity and $\Phi$-negativity, respectively, lead to two
sets of bounds on the entanglement of formation, the tangle, and the
concurrence. We compare these bounds and identify the sets of $4
\times N$ density operators for which the bounds from one constraint
are better than the bounds from the other. In the process, we present
a new derivation of the already known bound on the concurrence based
on the negativity.  We compute new bounds on the three measures of
entanglement using both the constraints simultaneously. We demonstrate how such
doubly constrained bounds can be constructed. We discuss extensions of our
results to bipartite states of higher dimensions and with more than two
constraints.

\end{abstract}
  


\maketitle

\section{Introduction}\label{S:intro}

Characterizing quantum entanglement~\cite{Schrodinger1935},
\cite{Bruss2002} is an important open problem in quantum
information theory~\cite{Nielsen2000}. The nonclassical
correlations associated with entanglement have been of immense
interest since the very inception of quantum mechanics
\cite{Einstein1935}, \cite{Bell1964}. Quantum information science
has identified entanglement as a potential resource. The ability
of quantum computers to solve classically hard problems
efficiently, the increased security of quantum cryptographic
protocols, the enhanced capacity of quantum channels---all these
are attributed to entanglement~\cite{Nielsen2000}. The presence of
entanglement has been related to quantum phase transitions and the
behavior of condensed systems~\cite{Ghosh2003},
\cite{Osborne2002}, \cite{Osterloh2002}. Entanglement has also
allowed the understanding of techniques such as
density-matrix-renormalization group in a new
light~\cite{Vidal2003}. A significant part of recent research in
theoretical quantum information science has centered around
understanding and characterizing entanglement. In spite of this,
entanglement remains a poorly understood feature of quantum
systems.

Although many tests have been devised which attempt to decide whether
a general quantum state is separable or not, this problem is known to
be NP-Hard~\cite{Gurvits2003}. Quantifying entanglement involves
devising functions acting on quantum states that, in some reasonable
way, order entangled states according to the degree of nonclassical
correlation possessed by them. Measures of entanglement can be
broadly divided into two classes depending on whether an efficient
way of computing them for arbitrary states exists or not. Tests for
separability can also be classified in a similar
fashion~\cite{Bruss2002}. {\em Computationally operational} measures of
entanglement are easy to calculate for any state, while there is no
known procedure for efficiently calculating {\em computationally nonoperational}
measures for an arbitrary state. From here on we abbreviate the descriptions
computationally operational and computationally nonoperational to
simply operational and nonoperational, respectively. Several
physically significant measures of entanglement are of the
nonoperational variety. This makes it important to place bounds on
the values of such measures. In this paper, we investigate the
problem of placing lower bounds on nonoperational measures of
entanglement for a quantum state assuming that we know the values of
one or more operational measures for that state.

The outline of this paper is as follows.  In Sec.~\ref{general} we start
with examples of both operational and nonoperational measures of
entanglement. We then discuss the general scheme of placing bounds on
nonoperational measures using operational ones as constraints. In
Sec.~\ref{sec:Phimap} we start with the separability criterion due to
Breuer~\cite{Breuer2006} and then show that a new, operational entanglement
measure, called the $\Phi$-negativity, can be extracted from it.  In
Sec.~\ref{singly} we use the $\Phi$-negativity to bound three
nonoperational measures of entanglement for $4\times N$ systems, namely,
the entanglement of formation, the tangle, and the concurrence. We compare
our results to the bounds based on another operational measure, the
negativity. In the process, we present a different way of deriving the
results in~\cite{Chen2005}. In Sec.~\ref{doubly} we obtain bounds on the
three nonoperational measures using both the negativity and
$\Phi$-negativity simultaneously as constraints. We also discuss how our
new bounds relate to previously known bounds in this section. Our
conclusions and future prospects are summarized in Sec.~\ref{conclusion}.

\section{General considerations}\label{general}

\subsection{Operational and nonoperational measures of entanglement}

A commonly used measure of entanglement for a pure-state $\ket{\Psi}$ of two
systems $A$ and $B$ is the entropy of the reduced density operator $\rho_A$ (or
$\rho_B$),
\begin{equation}
\label{vonneumann}
S(\rho_A)=-\tr(\rho_A \log\rho_A)=S(\rho_B)=-\tr(\rho_B \log\rho_B).
\end{equation}
We write this entropy either as a function $h(\Psi)$ of the state $\ket{\Psi}$
or as a function $H(\vec{\mu})$ of the vector of Schmidt coefficients of
$\ket{\Psi}$. It is a physically motivated quantity, in that it gives the rate
at which copies of a pure state can be converted, by using only local operations
and classical communication (LOCC), into copies of maximally entangled states
and vice versa~\cite{Bennett1996}. This measure can be elevated so that it
applies to bipartite mixed states also by taking the so-called convex-roof extension of
Eq.~(\ref{vonneumann}). This extended quantity is the entanglement of formation (EOF),
and it is defined as
\begin{equation}
\label{eof}
h(\rho)\equiv \min_{\{p_j, \ket{\Psi_j}\}}\bigg\{\sum_j p_j
h(\Psi_j)\biggr|\rho=\sum_jp_j \ket{\Psi_j}\bra{\Psi_j} \bigg\}.
 \end{equation}
The EOF provides an upper bound on the rate at which maximally
entangled states can be distilled from $\rho$ and a lower bound on
the rate at which maximally entangled states must be supplied to
create copies of $\rho$~\cite{Hayden2001}. Exact expressions for
the EOF of several classes of states are known. One of the
earliest, and simplest, was for an arbitrary state of two qubits
\cite{Wootters1998}. The EOF in that case, was presented in terms
of the concurrence, a subsidiary quantity. The concurrence itself
has since been identified as an entanglement monotone and extended
to higher-dimensional systems~\cite{Rungta2001},\cite{Rungta2003}.

The EOF and the concurrence are examples of a more general framework of defining
entanglement measures. Suppose we have an entanglement measure $g$ defined only
on pure states $\ket{\Psi}$, which is a concave function $G$ of Schmidt
coefficients $\vec{\mu}$ of the marginal density operator of $\ket{\Psi}$.  That
is, suppose $g$ has the form $g(\Psi)= G(\vec{\mu})$ on pure states.  This can
be extended to a measure on mixed states via the convex-roof extension,
 \begin{equation}
\label{monotone}
g(\rho)=\min_{\{p_j, \ket{\Psi_j}\}}\bigg\{\sum_j p_j
g(\Psi_j)\biggr|\rho=\sum_j p_j\ket{\Psi_j}\bra{\Psi_j} \bigg\}.
 \end{equation}
It has been proven~\cite{Vidal2000} that any $g(\rho)$ constructed in this
way is, on average, nonincreasing under LOCCs. An entanglement measure with this property is known
as an entanglement monotone. Besides the EOF and concurrence, other
examples of entanglement monotones include the tangle, relative entropy,
entanglement of distillation, etc. Each has its use in particular physical
contexts.  All the entanglement measures just mentioned have one feature
in common: they are nonoperational. The bottleneck in evaluating most of
these measures for mixed states is the minimization over all pure-state
decompositions. As a consequence, placing lower bounds on these measures
of entanglement for arbitrary states becomes important.

An alternate approach to detecting and quantifying entanglement is
based on the application of positive (but not completely positive)
maps on density operators
\cite{Stinespring1955},\cite{Stormer1963}, \cite{Choi1972},
\cite{Choi1974}, \cite{Choi1975}, \cite{Horodecki1996},
\cite{Terhal2001}, \cite{Rudolph2000}, \cite{Rudolph2002},
\cite{Chen2003}. In particular, a quantum state is separable if
and only if it remains positive semidefinite under the action of
{\em any}\/ positive map. Given a positive map, we can construct
an entanglement measure based on the spectrum of the density
operators under the action of the map \cite{Plenio2007},
\cite{Vidal2002}. Such measures are typically much easier to
calculate for general quantum states than the ones discussed
earlier because they do not involve the convex-roof construction.
Measures of entanglement based on positive maps are therefore
operational in nature. The negativity~\cite{Zyczkowski1998},
\cite{Vidal2002} is an example of an entanglement measure of this
sort, derived from the transpose map~\cite{Peres1996},
\cite{Horodecki1996}.

We can use the operational measures of entanglement as constraints
to obtain bounds on nonoperational, convex-roof-extended ones. The
complexity of the minimization in Eq.~(\ref{monotone}) is reduced
by solving it over a constrained set, instead of over all
pure-state decompositions. This was done in~\cite{Chen2005a},
\cite{Chen2005} for the EOF and the concurrence by minimizing over
states with a given value of negativity.  We turn now to
describing the general procedure for constructing bounds based on
the use of one or more operational entanglement measures as
constraints.

\subsection{Multiply-constrained bounds on nonoperational measures of
entanglement}\label{subsec:multiply}

Let $f_1,\cdots,f_K$ be $K$ operational measures used to characterize the
entanglement in a bipartite system. Assume that they have values ${\bf n} \equiv
(n_1,\ldots,n_K)$ for a state $\rho$. Their action on pure states can be
expressed as functions of the Schmidt coefficients, i.e.,
\begin{equation}
\label{eq:constraints} f_i(\Psi) =F_i(\vec {\mu}),\;\; i =
{1,\cdots,K}.
\end{equation}
We are interested in a lower bound on the value of an independent,
nonoperational measure $g$, which is a monotone defined on mixed
states via the convex-roof construction.  Let us assume that for the
state $\rho$, the optimal pure-state decomposition with respect to
$g$ is $\rho = \sum_j p_j |\Psi^j \rangle \langle \Psi^j|$. Then
\begin{equation}
\label{eq:doublyA2}
g(\rho) = \sum_j p_j \, g\big( \Psi^j \big) = \sum_j p_j \, G \big(\vec{\mu}^j
\big).
\end{equation}
Now define the function
\begin{equation}
\label{eq:doublyA3}
 \widetilde{G}(m_1,\ldots,m_K) \equiv \widetilde{G}({\bf m})=\min_{\vec
{\mu}}\big\{G(\vec {\mu}) \big| F_1(\vec{\mu})=m_1,\ldots, F_K(\vec{\mu})=m_K
\big\}.
\end{equation}
Let ${\mathcal G}({\bf m}) = {\mbox{co}} \bigl[ \widetilde{G} ({\bf m}) \bigr]$
be the convex hull of $\widetilde{G}({\bf m})$, i.e., the largest convex
function of $K$ variables $(m_1, \ldots , m_K)$ that is bounded from above by
$\widetilde{G}({\bf m})$. Using Eq.~(\ref{eq:doublyA3}) and the convexity of
${\mathcal G}$, we can write
\begin{equation}
\label{eq:doublyA5}
g(\rho)  \geq \sum_j p_j \, {\mathcal G}\big( {\bf n}^j \big) \geq {\mathcal G}
\biggl( \sum_j p_j {\bf n}^j \biggr).
\end{equation}
If ${\mathcal G}$ is a monotonically nondecreasing function of all its
arguments and if the operational measures $F_i$ are convex functions so that
$\sum_j p_j n^j_i \geq n_i$, we obtain
\begin{equation}
\label{eq:doublyA6}
g(\rho)  \geq {\mathcal G} ({\bf n}).
\end{equation}
If the conditions for the validity of the inequality~(\ref{eq:doublyA6}) are
met, then we obtain a lower bound on $g(\rho)$ by knowing the operational
measures ${\bf n}$ for $\rho$.

Regrettably, the first assumption leading to inequality
(\ref{eq:doublyA6}) is not always valid: the function
$\mathcal{G}({\bf n})$ is not guaranteed to be monotonic. If it is
not, then we have to impose monotonicity by introducing a new
monotonically nondecreasing function $\widetilde G_{\uparrow}({\bf
n})$, constructed from $\widetilde{G}({\bf n})$. In the examples we
consider in Sec.~\ref{doubly}, $\widetilde{G}({\bf n})$ turns out to
be monotonic, so we do not have to construct the new function
$\widetilde G_{\uparrow}({\bf n})$. For the sake of completeness, the
general construction of $\widetilde G_{\uparrow}({\bf n})$ is
presented in Appendix~{\ref{A:construction}}.

We can  now redefine ${\mathcal G}({\bf n})$ as the convex hull of
$\widetilde G_{\uparrow}({\bf n})$, rather than simply the convex
hull of $\widetilde G({\bf n})$. It is not immediately obvious that
the convex hull of a monotonically nondecreasing function is also
monotonically nondecreasing. The proof that this is so is given in
Appendix~\ref{AppProof}.

The only requirement on the operational entanglement
measures $F_i$ for using them as constraints is that they are convex
functions on the set of states.  Furthermore, even if we do not have
the functions $F_i$ themselves, but have instead functions
$\hat{F}_i$ that bound $F_i$ from above for pure states, then the
functions $\hat{F}_i$ can be used as constraints in the
definition~(\ref{eq:doublyA3}) of $\widetilde G$, in place of the
functions $F_i$.  The arguments leading to
inequality~(\ref{eq:doublyA6}) go through exactly as before, i.e.,
$g(\rho)\geq\sum_j p_j{\mathcal G}(\hat{\bf n}^j)\geq{\mathcal G}
\Bigl(\sum_j p_j \hat{\bf n}^j\Bigr)\geq\mathcal{G}\Bigl(\sum_j
p_j{\bf n}^j\Bigr)\geq{\mathcal G}({\bf n})$, the only difference
being that there is an additional step, the second-to-last one, where
we use $\hat{F}_i (\vec{\mu}) = \hat {n}_i \geq n_i = F_i$ to
conclude that $\sum_j p_j \hat n^j_i \geq \sum_j p_j n_j$.  The
danger in using upper bounds instead of the actual values of the
functions $F_i$ is that the final bound on $g(\rho)$ might turn out
to be less useful or even meaningless.  In the example we consider in
Sec.~\ref{doubly}, however, one of the constraints we use is an upper
bound on an operational entanglement measure, rather than the measure
itself, yet the bound we get turns out to be stronger than previous
bounds.

Since our bound is intended for arbitrary states, there is one more
subtlety to address, and that is the domain of the functions
$G,\widetilde{G}$, and $\mathcal{G}$.  The
operational measures ${\bf n}$ map the state $\rho$ to a point in a
$K$-dimensional hypercube in the space of the $K$ independent
constraints $n_k$.  Pure states correspond to a simply connected
subset in this hypercube, which we call the pure-state region. The
pure-state region is the domain of the functions $G$ and  $\widetilde{G}$. This
domain is not always convex, and so $\mathcal{G}({\bf n})$ is defined on the
convex hull of the pure-state region, which is generally bigger than the
pure-state region, though only a subset of the full hypercube available to a
general state.

Finally, we have to extend ${\mathcal G}({\bf n})$ to the entire hypercube of
states. Note that for inequalities (\ref{eq:doublyA5}) and (\ref{eq:doublyA6})
to hold, ${\mathcal G}({\bf n})$ must be a monotonically nondecreasing function
in the entire hypercube while it has to be convex only on the convex hull
of the pure-state region. So, in extending ${\mathcal G}({\bf n})$ outside the
hull, we only have to take into account the monotonicity requirement
(\ref{eq:doublyA6}). To construct such an extension of ${\mathcal G}({\bf n})$,
start from a point on the boundary of the hull and begin traversing out along
{\em decreasing} directions parallel to the axes of the hypercube. Outside the
hull, and till reaching the boundaries of the hypercube, the extension is
defined as the constant function with value equal to that at the point on the
boundary of the hull. To generate the complete extension, this simple procedure
is repeated for every point on all the boundaries of the hull. This procedure
is also demonstrated in detail in Sec.~\ref{doubly} for the example we consider.

In this paper, we carry out the general program just described
with two particular constraints ($K=2$). One of them is the
negativity \cite{Vidal2002}. For the second, we develop a new
entanglement measure, called $\Phi$-negativity, based on a
recently presented separability criterion \cite{Breuer2006}
(see~\cite{Breuer2006a} for another measure based on the same
criterion). Like the negativity, it is easily computable for any
$\rho$ and there are no convex-roof constructions involved in the
computation. The $\Phi$-negativity, unlike the negativity, is not
a simple function of the Schmidt coefficients for pure states. We
find a simple function of the Schmidt coefficients that is an
upper bound on the $\Phi$-negativity and, as described above, we
use this function instead as the constraint to simplify our
computations.  We use both the (upper bound on) $\Phi$-negativity
and the negativity simultaneously as constraints to place new
bounds on the EOF, tangle, and concurrence of $4\times N$ systems.
Ours is the first instance of a doubly-constrained bound on
entanglement measures for a family of states. It puts bounds that
are tighter than those obtained in~\cite{Chen2005a},
\cite{Chen2005}. Multiply constrained bounds based on entanglement
witnesses that can be applied to individual quantum states have
been obtained using a different approach in \cite{Guhne2006},
\cite{Eisert2006}.

Although all of the results in this paper are obtained using the
negativity and $\Phi$-negativity, a third constraint based on the
realignment criterion~\cite{Rudolph2000}, \cite{Rudolph2002},
\cite{Chen2003} can be added to improve the bounds for certain
classes of states. On pure states, the negativity and the
realignment criterion lead to the same constraint. This means that
in deriving both the singly and doubly constrained bounds we could
have modified the negativity to take advantage of this, as was
done in~\cite{Chen2005a}, \cite{Breuer2006a}.  Furthermore, the
addition of the realignment criterion adds very little complexity
to the procedure described below.

Before concluding this section, we review the notation used in this
paper. We use lower case Latin letters, say $g$, to denote
entanglement measures. The corresponding upper case character, $G$,
denotes the same entanglement measure defined on pure states,
expressed as a function of the Schmidt coefficients. The same letter
with a tilde on top, $\widetilde{G}$, stands for the minimum of $G$
subject to constraints. Calligraphic letters like ${\mathcal G}$
denote the bound on $g$ obtained by taking the convex hull of
$\widetilde{G}$. If we have to impose monotonicity on $\widetilde{G}$
as an intermediate step, we define a new
function~$\widetilde{G}_{\uparrow}$.

\section{\texorpdfstring{$\Phi$-}{Phi-}Map}\label{sec:Phimap}

Recently, a new separability criterion has been proposed based on
a positive nondecomposable map \cite{Breuer2006}. It is a
combination of the Peres criterion and the reduction
criterion~\cite{Cerf1999}, \cite{Horodecki1999} for detecting
entangled states. In this section we construct a new entanglement
measure from this map and calculate it for pure states.

\subsection{Separability criterion}

Let us consider a finite-dimensional Hilbert space $\mathbb{C}^D$. It
can be regarded as the space of a spin-$j$ particle with $D=2j+1$. A
natural basis for this space is the ``angular-momentum basis''
$\ket{j,m}$, where $m=-j, -j +1, \dots, j-1, j$. The separability
criterion to be presented involves the time-reversal operator
$\vartheta$ whose action on an operator $\sigma$ acting on
$\mathbb{C}^D$ is given as
\begin{equation}
\vartheta \sigma = V \sigma^T V^{\dag},
\end{equation}
where the superscript $T$ stands for transposition in the
angular-momentum basis and $V$ is a unitary operator defined as
\begin{equation}
\label{eq:vmat}
\bra{j,m}V\ket{j,m'}= (-1)^{j-m}\delta_{m,-m'}.
\end{equation}
This map was initially introduced by Breuer to study the entanglement of
$4 \times 4 \; \mathrm{SU}(2)$ invariant states; in that case, the
$\vartheta$ map, together with the Peres criterion, was found to be a
necessary and sufficient separability condition~\cite{Breuer2005}. In even
dimensions, an additional property holds: $V^T=-V$, i.e., $V$ is
skew-symmetric in addition to being unitary.

The condition for positivity under the partial time-reversal map $(I
\otimes \vartheta)\rho \geq 0$ is unitarily equivalent to the Peres PPT
criterion $(I \otimes T)\rho \geq 0$. This means that partial time
reversal can be used as an entanglement detection criterion.   Breuer
\cite{Breuer2006} defines a positive map
\begin{equation}
\label{phimap}
\Phi(\rho) = \tr(\rho)I -\rho - V\rho^T V^{\dag}\;,
\end{equation}
which conjoins the time reversal map with the so-called reduction
criterion~\cite{Cerf1999}. The map $\Phi$ then defines for any joint density
operator $\rho_{AB}$ a necessary condition for separability as
 \begin{equation}
 \label{totalphi} (I\otimes\Phi)\rho_{AB} = \tr_{B}(\rho_{AB})\otimes
I_B-\rho_{AB} - (I_A\otimes V)\rho_{AB}^{T_B}(I_A\otimes
V^{\dag})\geq 0.
 \end{equation}
Any state that violates the above condition must be entangled.

Consider the space $\mathcal{H}_A\otimes
\mathcal{H}_B=\mathbb{C}^D \otimes \mathbb{C}^D$. It can be
regarded, without loss of generality, as the Hilbert space of two
spin-$j$ particles with $j=(D-1)/2$. The total spin of the system,
denoted by $J$ ranges over the values $J=0,1,\dots,2j=D-1$. Let
$P_J$ be the projector onto the ($2J+1$)-dimensional spin-$J$
manifold. It can then be shown that $\Phi$ is a nondecomposable
positive, but not completely positive map~\cite{Breuer2006},
\cite{Breuer2006a} in all even dimensions $D$ greater than or
equal to 4.  The proof of positivity cannot be extended to odd
dimensions as it exploits the skew-symmetric nature of the unitary
operator $V$. In addition, the hermitian operator
 \begin{equation}
 W \equiv (I\otimes \Phi)P_0
 \end{equation}
is an optimal entanglement witness~\cite{Lewenstein2000},
\cite{Breuer2006}, in that the set of PPT states detected by $W$
is not contained in the set detected by any other \emph{single}
witness. There, of course, exist families of PPT states that $W$
fails to detect. The optimal nature of $W$ provides motivation for
contructing a measure of entanglement based on the $\Phi$-map.

\subsection{Entanglement measures from maps}\label{sec2B}

Our endeavor here is to define a quantitative operational
measure of entanglement based on the $\Phi$ map. We call
this quantity the $\Phi$-negativity, denote it by $n_{\Phi}$, and
define it for a general mixed state as
 \begin{equation}
 \label{nphi}
n_{\Phi}(\rho)=
\frac{D(D-1)}{4}\left[\frac{||(I\otimes\Phi)\rho||}{D-2}-1\right],
 \end{equation}
where $D=\min(\dim(\mathcal{H}_A),\dim(\mathcal{H}_B))$, and the
trace norm of an operator is defined as
$||O||=\tr(\sqrt{OO^{\dag}})$. For a separable state $\sigma$,
$(I\otimes\Phi)\sigma$ has no negative eigenvalues, so
$||(I\otimes\Phi)\sigma||= \tr{[(I\otimes\Phi) \sigma]} = (D-2)
\tr(\sigma)=D-2.$ Hence the $\Phi$-negativity is zero on separable
states. This calculation also shows that $\Phi$ is not trace
preserving, and this motivates the factor of $D-2$ in the denominator
of Eq.~(\ref{nphi}). The $\Phi$-negativity is a shifted and scaled
version of the sum of the negative eigenvalues of a state under the
action of the map~(\ref{totalphi}). Since this sum can be
expressed in terms of the trace norm of an operator,
$||(I\otimes\Phi)\rho_{AB} ||$, it is a convex function of $\rho$ as
required in the general scheme described in Sec.~\ref{general}. By
defining the $\Phi$-negativity in terms of a map, we make sure that
it is an operational measure that involves no convex-roof extensions.

Similar measures of entanglement based on other positive, but not
completely positive maps have previously been proposed and
investigated~\cite{Plenio2007}. The negativity~\cite{Vidal2002}, which is based on
the Peres partial transpose criterion, is defined as
 \begin{equation}
 n_T(\rho) = \frac{||\rho^{T_A}||-1}{2},
 \end{equation}
where $\rho$ is a joint density operator, $T_A$ is the partial
transposition with respect to system $A$. A positive value of $n_T$
indicates an entangled state. The Peres negativity, in addition to
being a measure of entanglement, is also an entanglement monotone,
since it is nonincreasing on average under LOCC
operations~\cite{Vidal2002}. The $\Phi$-negativity is not an
entanglement monotone, but it is a convex function of $\rho$.

The $\Phi$-negativity is a new operational measure of entanglement
for any quantum state.  To use the $\Phi$-negativity as a constraint
in bounding nonoperational measures we need expressions for
$n_{\Phi}$ for pure states.  We start from the Schmidt decomposition
of any pure state,
 \begin{equation}
 \label{schmidt}
 \ket{\Psi_{AB}}=\sum_{i=1}^{D} \sqrt{\mu_i}\ket{a_i,b_i}
 \end{equation}
for $\ket{\Psi_{AB}}\in \mathbb{C}^D\otimes\mathbb{C}^N$ and $D\leq
N$. The $\mu_i$ are the Schmidt coefficients, satisfying $\mu_i \geq
0 \;\; \forall\;i$ and $\sum_{i=1}^D \mu_{i}=1$.

Before we apply the map $(I \otimes \Phi)$ to this state we note that
the matrix $V$ appearing in the definition of the $\Phi$-map has the
form given in Eq.~(\ref{eq:vmat}) only in the angular-momentum basis
for system $B$, and the required transposition is also carried out in
this basis.  Relabeling the angular-momentum eigenvectors
$\{\ket{j,m} \}$ as $\{ \ket{l} \}$ with $l=1,\ldots,D=2j+1$, we
transform $|\Psi_{AB} \rangle$ to the angular-momentum basis for
subsystem $B$ and obtain
\begin{equation}
\rho_{AB} =\ket{\Psi_{AB}}\bra{\Psi_{AB}}= \sum_{i,j,l,m=1}^D\sqrt{\mu_i \mu_j}
\langle l | b_i \rangle \langle b_j | m \rangle |a_i, l \rangle \langle a_j,
m|,
 \end{equation}
and
 \begin{eqnarray}
 \label{phionrho1}
(I\otimes \Phi)\rho_{AB} & = & \sum_{i=1}^D \mu_i\ket{a_i}\bra{a_i} \otimes
\sum_{l=1}^D \ket{l}\bra{l} \nonumber \\
&& -\sum_{i,j,l,m=1}^D\sqrt{\mu_i \mu_j}
\langle l | b_i \rangle \langle b_j | m \rangle |a_i, l \rangle \langle a_j,
m| \nonumber \\
&&  -\sum_{i,j,l,m=1}^D \sqrt{\mu_i \mu_j}
(-1)^{l+m} \langle b_i | l \rangle \langle m | b_j \rangle
\ket{a_i,D+1-m}\bra{a_j, D+1-l}.
 \end{eqnarray}
The trace norm $||(I \otimes \Phi)\rho_{AB}||$ and the entanglement
measure $n_{\Phi}$ defined using the trace norm are rather
complicated functions of the Schmidt coefficients $\mu_i$ and the
matrix elements $\langle l| b_i\rangle$ of the unitary matrix that
transforms between the Schmidt basis of subsystem $B$ the angular
momentum basis used in Eq.~(\ref{eq:vmat}). Computing the numerical
value of $n_{\Phi}$ for any state is relatively easy, but the
analytic expression for the entanglement measure is quite unwieldy.

All we really need to generate a constraint from the $\Phi$-map,
which can be used to place lower bounds on nonoperational measures of
entanglement, is an upper bound on $n_{\Phi}$. Such a bound is
obtained by considering the special case in which the Schmidt basis
of subsystem $B$ is the same as the basis of the angular-momentum
eigenstates $\{ \ket{l} \}$, i.e., $\langle l | b_i \rangle =
\delta_{li}$. We then have
 \begin{eqnarray}
 \label{phionrho}
(I\otimes \Phi)\rho_{AB} & = & \sum_{i=1}^D \mu_i \ket{a_i}\bra{a_i} \otimes
\sum_{j=1}^D \ket{b_j}\bra{b_j} \nonumber \\
&& -\sum_{i,j=1}^D \sqrt{\mu_i\mu_j}\Bigl[\ket{a_i b_i}\bra{a_j
b_j}+(-1)^{i+j}\ket{a_i}\bra{a_j}\otimes\ket{b_{D-j+1}}\bra{b_{D-i+1}}\Bigr].
 \end{eqnarray}
For the first nontrivial case, $D=4$, which we will be using
extensively, explicit diagonalization of the above operator is
possible.  As shown in Appendix~\ref{appA}, it has six nonzero
eigenvalues, one of which is negative. The trace norm can then be
evaluated as the sum of the absolute values of the eigenvalues. Thus,
\begin{equation}
\label{tnorm}
 ||(I \otimes \Phi)\rho_{AB}||= \tr\Bigl(\sqrt{[(I \otimes
\Phi)\rho_{AB}]^2}\Bigr) = 2[1+\sqrt{(\mu_1+\mu_4)(\mu_2+\mu_3)}],
\end{equation}
where we use the fact that the $\Phi$-map is hermiticity preserving.
Therefore, for all $4\times N$ pure states that have the Schmidt
basis for subsystem $B$ the same as the angular-momentum basis,
 \begin{equation}
 \label{nphi4}
 n_{\Phi}=3\sqrt{(\mu_1+\mu_4)(\mu_2+\mu_3)} \equiv \hat{n}_{\Phi}.
 \end{equation}
In Eqs.~(\ref{tnorm}) and (\ref{nphi4}) and in all our subsequent
discussion of the function $\hat n_\Phi$, the Schmidt coefficients
are ordered from largest to smallest, i.e.,
$\mu_1\ge\mu_2\ge\mu_3\ge\mu_4$.

\begin{figure}[!ht]
\resizebox{8.1cm}{5.5cm}{\includegraphics{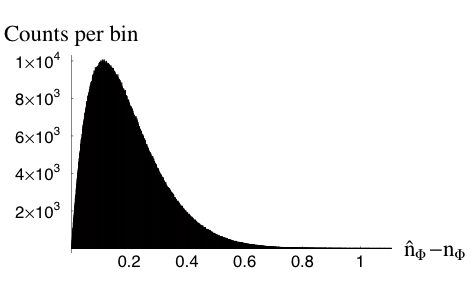}}
\caption{A histogram of the difference $\hat{n}_{\Phi} - n_{\Phi}$
for five million $4 \times 4$ pure states picked randomly from the
Haar measure. The bin size in the histogram is $0.001$. The
difference is always found to be positive.  These results carry over
to $4 \times N$ states when $N>4$ ($N$ even), since the difference
only depends on the Schmidt coefficients.  We have tried to prove
that the difference is nonnegative without success and thus rely on
this numerical demonstration instead.}
 \label{bound}
\end{figure}

The function $\hat{n}_{\Phi}$ is a simple function of the Schmidt
coefficients for any $4 \times N$ pure state and as shown numerically
by the results displayed in Fig.~\ref{bound}, the true
$\Phi$-negativity, $n_{\Phi}$, calculated with respect to a fixed
angular-momentum basis, is always bounded from above by
$\hat{n}_{\phi}$.

In the rest of this paper we use $\hat{n}_{\Phi}$ instead of
$n_{\Phi}$ as the constraint for bounding nonoperational measures
because of its simple algebraic form.  When we refer to constraints
based on the $\Phi$-negativity we are referring to fixing the value
of $\hat{n}_{\Phi}$. Expressions for $\hat{n}_{\Phi}$ for pure states
in higher dimensions are discussed in Appendix~\ref{appA}.

In the next section, we will use $\hat{n}_{\Phi}$ to put lower
bounds on the EOF, tangle, and concurrence for $4 \times N$ mixed
states. We then compare our results to such bounds that have
already been derived based on the Peres negativity $n_T$. For $D
\times N$ pure states, the Peres negativity is given
by~\cite{Zyczkowski1998},\cite{Vidal2002}
 \begin{equation}
 n_T = \frac{\left(\sum_{i=1}^D\sqrt{\mu_i} \right)^2-1}{2}.
 \end{equation}

\section{Singly Constrained Bounds}\label{singly}

\subsection{Entanglement of formation}\label{singly:eof}

A lower bound ${\mathcal H}\big( \hat{n}_{\Phi} \big)$ on the EOF, constrained
by pure states having a certain value for $\hat{n}_{\Phi}$,
can be obtained using the steps described in Sec.~\ref{general}.
All the subsequent results presented in this section and the
next are for $4\times N$ states $\rho$, with $N\ge4$.

Firstly, we have to find
  \begin{equation}
\widetilde{H}\big( \hat{n}_{\Phi}
\big)=\min_{\vec{\mu}}\left\{H(\vec{\mu})\Bigl|3\sqrt{(\mu_1 + \mu_4)(\mu_2 +
\mu_3)}=\hat{n}_{\Phi} \right\},
  \end{equation}
and then its convex hull,
 \begin{equation}
 {\mathcal H }(\rho) = \mathrm{co}\big[\widetilde{H}\big( \hat{n}_{\Phi}
\big)\big],
 \end{equation}
provided $\widetilde{H} \big( \hat{n}_{\Phi} \big)$ is a monotonically
increasing
function of $\hat{n}_{\Phi}$.  Defining $\mu_1+\mu_4=\alpha$ and
$\mu_2+\mu_3=\beta$, we can write the normalization and $\hat{n}_{\Phi}$
constraints as
 \begin{equation}
 \label{cons}
 \alpha + \beta = 1\qquad\mbox{and}\qquad\alpha\beta =
\frac{\hat{n}_{\Phi}^2}{9},
\end{equation}
which give
 \begin{equation}
 \label{phialpha}
 \alpha=\frac{1\pm\sqrt{1-4\hat{n}_{\Phi}^2/9}}{2}
 \quad\mbox{and}\quad
 \beta=\frac{1\mp\sqrt{1-4\hat{n}_{\Phi}^2/9}}{2}
 \end{equation}
Minimizing
\begin{equation}
  H(\vec\mu)=-\mu_1\log\mu_1-\mu_4\log\mu_4
  -\mu_2\log\mu_2-\mu_3\log\mu_3=
  H_2(\alpha)+\alpha H_2(\mu_1/\alpha)+\beta H_2(\mu_2/\beta),
\end{equation}
where $H_2(\cdot)$ is the binary entropy function, is trivial,
because we simply make the last two terms zero by choosing
$\mu_1=\alpha$ and $\mu_2=\beta$ [and choosing the upper sign in
Eq.~(\ref{phialpha}) so as to be consistent with the assumed ordering
of the Schmidt coefficients]. Then the minimum entropy is
 \begin{equation}
 \widetilde{H}\big( \hat{n}_{\Phi} \big) = H_2(\alpha).
 \end{equation}

\begin{figure}[!ht]
\resizebox{8.1cm}{6cm}{\includegraphics{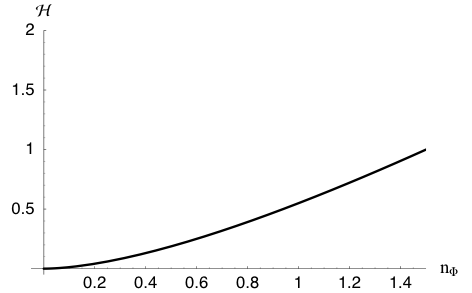}}
\resizebox{8.1cm}{6cm}{\includegraphics{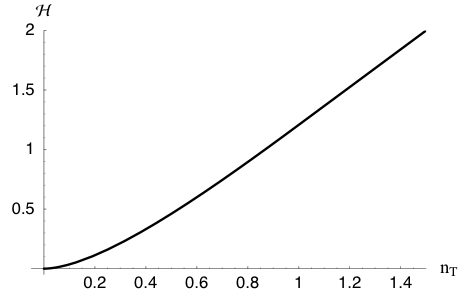}}
 \caption{On the left is the bound on the EOF based on a constrained
$\hat{n}_{\Phi}$, Eq.~(\ref{phibound}). The plot on the right is the
bound on the EOF based on a constrained negativity,
Eq.~(\ref{Tbound}).}
 \label{Rtphifig}
\end{figure}

That $\widetilde{H}\big( \hat{n}_{\Phi} \big)$ is a convex, monotonically
increasing function of $n_\Phi$  can be shown by considering its first and
second derivatives. Its convex roof is the function itself, i.e.,
\begin{equation}
{\mathcal H}\big( \hat{n}_{\Phi} \big)
=\mathrm{co}\big[\widetilde{H}\big(
\hat{n}_{\Phi} \big)\big]=\widetilde{H}\big( \hat{n}_{\Phi} \big),
\end{equation}
and the bound can thus be extended to mixed states, giving
 \begin{equation}
 \label{phibound}
 h(\rho) \geq H_2\!\left(\frac{1+\sqrt{1-4{n}_{\Phi}^2/9}}{2}\right),
 \end{equation}
with $n_\Phi$ being the $\Phi$-negativity of $\rho$.

The first step in bounding the EOF with only a single constraint on the
negativity is to determine the function
 \begin{equation}
 \label{negmin}
\widetilde{H}\big( n_T \big)
=\min_{\vec{\mu}}\left\{H(\vec{\mu})\left|\frac{\left(\sum_{j=1}^4
\sqrt{\mu_j}\right)^2-1}{2} = n_T \right.\right\}.
 \end{equation}
This was solved in~\cite{Terhal2000},\cite{Chen2005a} for $2$ or
$3$ Schmidt coefficients and recently shown to be valid for any
number of Schmidt coefficients~\cite{Fei2006}.  In particular, for
four Schmidt coefficients, the case of interest here, we obtain
 \begin{equation}
\widetilde{H}\big( n_T \big) = H_2(\gamma) + (1-\gamma)\log_2 3,
 \end{equation}
with
 \begin{equation}
\label{eq:gamma}
\gamma=\frac{\left(\sqrt{2n_T+1} + \sqrt{3(3-2n_T)}\right)^2}{16}.
 \end{equation}

\begin{figure}[!ht]
\resizebox{8 cm}{8
cm}{\includegraphics{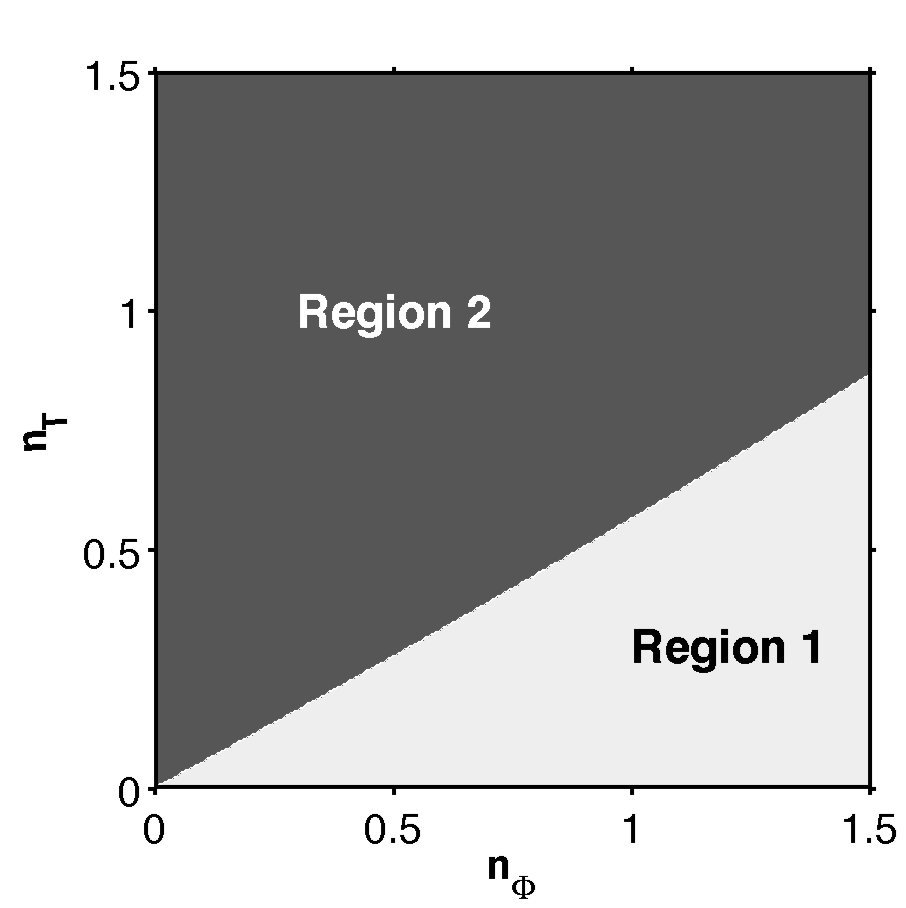}}
 \caption{In Region 1, the singly-constrained $n_{\Phi}$ bound is better
than the singly-constrained $n_T$ bound. In Region 2, the opposite is
true.}
 \label{singlecompareF}
\end{figure}

Unlike $\widetilde{H}\big( \hat{n}_{\Phi} \big)$, $\widetilde{H}\big( n_T \big)$
is not convex over the entire range of $n_T$. It is, however, a
monotonically increasing function of $n_T$. The actual bound on the EOF is
thus the convex-roof extension of this function,
$\mathrm{co}[\widetilde{H}\big( n_T \big)]$, which is given
as~\cite{Chen2005a}
 \begin{equation}
\label{Tbound}
\begin{array}{l}
h(\rho) \geq {\mathcal H}\big( n_T \big) \equiv
\mathrm{co}\big[\widetilde{H}\big( n_T \big)\big]=
\left\{ \begin{array}{ll}
 H_{2}(\gamma)+(1-\gamma)\log _{2} 3, & n_T \in [0,1],\\[2mm]
 \big(n_T -\frac{3}{2}\big)\log _2 3+ 2, & n_T \in [1,\frac{3}{2}].
\end{array} \right. \end{array}
\end{equation}

Both the singly constrained bounds are plotted in Fig.~\ref{Rtphifig}. It
might seem that the bound based on the $\hat{n}_{\Phi}$ constraint is
always poorer than that in Eq.~(\ref{Tbound}), but this is not the case.
There is a region in the $\hat{n}_{\Phi}$-$n_T$ plane where the bound of
Eq.~(\ref{phibound}) is better than that of Eq.~(\ref{Tbound}).  This is
depicted in Fig.~\ref{singlecompareF}.

\subsection{Tangle and concurrence}\label{singly:tangle}

The procedure in the previous section can be undertaken for the
tangle $t(\rho)$ and the concurrence $c(\rho)$ \cite{Rungta2001},
\cite{Rungta2003}. To place bounds on the tangle, we start by
finding
 \begin{equation}
\widetilde{T}\big( \hat{n}_{\Phi}
\big)=\min_{\vec{\mu}}\left\{2\left(1-|\vec{\mu}|^2\right) \Biggl|3\sqrt{(\mu_1
+ \mu_4)(\mu_2 + \mu_3)}=\hat{n}_{\Phi} \right\},
 \end{equation}
which gives a bound for pure states.  Then, just as for the EOF, the bound
on the tangle for mixed states is given by the convex hull of
$\widetilde{T}\big( \hat{n}_{\Phi} \big)$,
 \begin{equation}
t (\rho) \geq  {\mathcal T}\big( \hat{n}_{\Phi} \big)  \equiv
\mathrm{co}\big[\widetilde{T}\big( \hat{n}_{\Phi} \big)\big],
 \end{equation}
provided $\widetilde{T}\big(\hat{n}_{\Phi} \big)$ is a monotonically
nondecreasing
function of $\hat{n}_{\Phi}$.

Using the normalization and $\hat{n}_{\Phi}$ constraints of
Eq.~(\ref{phialpha}), we have
\begin{equation}
2\left(1-|\vec{\mu}|^2\right)=2\left(1-\sum_{i=1}^4\mu_i^2\right)=
4\sum_{i<j}\mu_i\mu_j
=4\left(\frac{\hat{n}_{\Phi}^2}{9}+\mu_1\mu_4+\mu_2\mu_3\right).
 \end{equation}
Just as for the EOF, the minimization is trivial, the minimum
occurring for the upper sign in Eq.~(\ref{phialpha}), with $\mu_4=0$
($\mu_1=\alpha$) and $\mu_3=0$ ($\mu_2=\beta$), thus giving
 \begin{equation}
\widetilde{T}\big( \hat{n}_{\Phi} \big)=\frac{4}{9}\hat{n}_{\Phi}^2.
 \end{equation}
Since this is both monotonically increasing and convex in
$\hat{n}_{\Phi}$, the same bound holds for mixed states, but in terms
of the actual negativity $n_\Phi$, i.e.,
 \begin{equation}
 \label{concnphibound}
t(\rho) \geq  {\mathcal T}\big({n}_{\Phi} \big) =
\frac{4}{9}{n}_{\Phi}^2.
 \end{equation}

The lower bound on the tangle, subject to a constraint on the negativity,
is found by starting from
 \begin{equation}
\widetilde{T} \big( n_T
\big)=\min_{\vec{\mu}}\left\{2\left(1-|\vec{\mu}|^2\right)\Biggl|\frac{
\left(\sum_{j=1}^4 \sqrt{\mu_j}\right)^2-1}{2}=n_T \right\}.
 \end{equation}
This is a relatively involved minimization, but it is exactly the same as
the minimization problem that arises in evaluating a bound on the tangle
for isotropic states, so we can adapt the result of~\cite{Rungta2003} to
give
\begin{equation}
\label{rungtacavesbound}
\widetilde{T} \big( n_T \big) =
\frac{1}{12}\left(9+4n_T^2+\sqrt{3\left(3+4n_T-4n_T^2\right)}(2n_T-3)
\right).
 \end{equation}
This quantity is monotonically increasing, but is not convex over the
complete range of $n_T$.  The convex hull ${\mathcal T}\big( n_T \big)
\equiv \mathrm{co}\big[\widetilde{T} \big( n_T \big)\big]$ is required to
extend the bound to mixed states.  Again using the results
of~\cite{Rungta2003}, we obtain
 \begin{equation}
 \label{concntbound}
\begin{array}{l}
t(\rho) \geq {\mathcal T}\big( n_T \big) = \left\{
\begin{array}{ll}
\frac{1}{12}\left(9+4n_T^2+\sqrt{3\left( 3+4n_T-4n_T^2\right)}(2n_T-3) \right),
& n_T \in [0,1],\\[2mm]
\frac{4}{3}n_T-\frac{1}{2}, & n_T \in [1,\frac{3}{2}]. \end{array}
\right. \end{array}
 \end{equation}

We can derive from Eq.~(\ref{concnphibound}) an expression for the
lower bound on the concurrence of $4 \times N$ states with a given
value of ${n}_{\Phi}$:
\begin{equation}
c(\rho) \geq {\mathcal C}\big( n_{\Phi} \big) = \widetilde{C} \big(
n_{\Phi}
\big) = \sqrt{\widetilde{T} \big( n_{\Phi} \big)} = \frac{2}{3}
n_{\Phi}.
\end{equation}
An expression for the minimum of the concurrence, subject to the
negativity constraint, can be obtained from Eq.~(\ref{rungtacavesbound}).
The resulting function is everywhere concave, and thus its convex hull is
a straight line joining the end points. This line is
 \begin{equation}
 \label{chenbound}
c(\rho) \geq \mathcal{C} \big( n_T \big) = \sqrt{\frac{2}{3}}n_{T}.
 \end{equation}

\begin{figure}[!ht]
\resizebox{8.1cm}{6cm}{\includegraphics{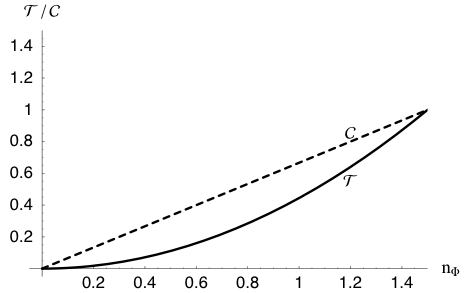}}
\resizebox{8.1cm}{6cm}{\includegraphics{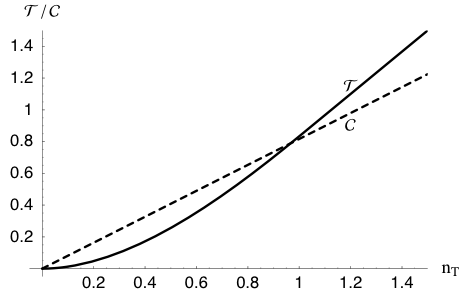}}
 \caption{The plot on the left shows the bounds on the tangle and the
concurrence based on the $n_{\Phi}$ constraint. The solid line is
the bound on the tangle and the dashed line is the bound on the
concurrence. On the right is a plot of the analogous bounds based on the
$n_T$ constraint.}
 \label{Ctphifig}
\end{figure}

\begin{figure}[!ht]
\resizebox{8 cm}{8 cm}{\includegraphics{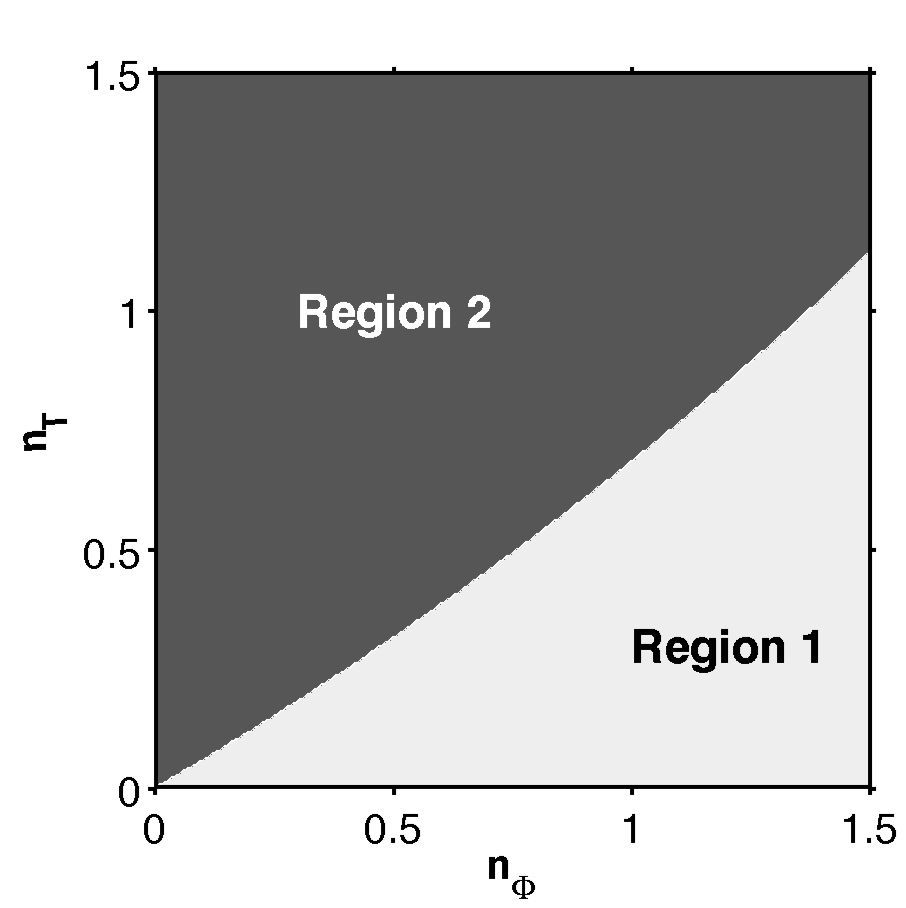}}
\caption{Region 1 is where the $n_{\Phi}$ constraint is better than the
$n_T$ constraint for bounding the tangle and concurrence.  Region 2 is where the
converse is true.}
 \label{singlecompareT}
\end{figure}

The bounds on both the tangle and the concurrence are plotted in Fig
\ref{Ctphifig}. As was true for the EOF, the $n_\Phi$ bound is better than the
$n_T$ bound in some parts of the ${n}_{\Phi}$-$n_T$ plane. This is shown in
Fig.~\ref{singlecompareT}.

Recently, a lower bound on the concurrence has been derived based on the
negativity constraint~\cite{Chen2005}, using techniques different from those
employed here. That lower bound is exactly the one in Eq.~(\ref{chenbound}). We
have thus provided an independent derivation of the bound presented
in~\cite{Chen2005}. In addition, we can use the procedure from \cite{Chen2005} to
derive a lower bound on the tangle based on the $\hat{n}_{\Phi}$ constraint.
Then we obtain  \begin{equation}
 \frac{\widetilde{T} \big( \hat{n}_{\Phi} \big)}{4} -
\frac{\hat{n}_{\Phi}^2}{9}=
\mu_1\mu_4+\mu_2\mu_3\geq 0,
 \end{equation}
which for general mixed states, leads exactly to the bound in
Eq.~(\ref{concnphibound}).

\section{Doubly Constrained Bounds} \label{doubly}

In this section we place new lower bounds on the EOF, tangle, and
concurrence for $4 \times N$ density operators by using $n_T$ and
$\hat{n}_{\Phi}$ simultaneously as constraints.

\subsection{Pure states of \texorpdfstring{$4 \times N$}{4 by N} systems}
\label{doubly:pure}

For a $4 \times N$ pure state, described by the Schmidt coefficients
$\mu_i$, $i=1,\ldots,4$, we have three constraint equations,
\begin{eqnarray}
\label{eq:doublyB1}
\frac{1}{2} \left[ \left( \sqrt{\mu_1} + \sqrt{\mu_2} + \sqrt{\mu_3} +
\sqrt{\mu_4}\,\right)^2 -1 \right] & = & n_T, \nonumber \\
3 \sqrt{(\mu_1 + \mu_4)(\mu_2 + \mu_3)} &=& \hat{n}_{\Phi}, \nonumber \\
\mu_1 + \mu_2 + \mu_3 + \mu_4 &=&1,
\end{eqnarray}
in addition to the inequality constraints
\begin{equation}
1\ge\mu_1\ge\mu_2\ge\mu_3\ge\mu_4\ge0.
\end{equation}
Both $\hat{n}_{\Phi}$ and $n_T$ take on values between $0$ and $3/2$,
so all $4 \times N$ states, pure or mixed, are mapped to a square of
side $3/2$ in the $\hat{n}_{\Phi}$-$n_T$ plane. Not all points in the
square correspond to pure states. If we solve the three equations in
(\ref{eq:doublyB1}) simultaneously and express $\mu_1$, $\mu_2$ and
$\mu_3$ in terms of $n_T$, $\hat{n}_{\Phi}$ and $\mu_4$ (see
Appendix~\ref{appB}), we find that for some allowed values of
$\hat{n}_{\Phi}$ and $n_T$, there is no allowed value of $\mu_4$ for
which the other three Schmidt coefficients are real numbers between
$0$ and $1$ in even one of the solution branches of
(\ref{eq:doublyB1}).

To find the region occupied by pure states in the $\hat{n}_{\Phi}$-$n_T$ plane,
let us use the pure-state expressions for $n_T$ and $\hat{n}_{\Phi}$ in
Eq.~(\ref{eq:doublyB1}) to find the largest and smallest values that $n_T$
can take on for a fixed value of $\hat{n}_{\Phi}$.  We proceed as in the
minimization of $H(\vec\mu)$ in Sec.~\ref{singly}.  Defining
$\alpha=\mu_1+\mu_4$ and $\beta=\mu_2+\mu_3$, the normalization and
$\hat{n}_{\Phi}$ constraints can be solved to give $\alpha$ and $\beta$ as in
Eq.~(\ref{phialpha}).  The negativity takes the form
\begin{equation}
\sqrt{2n_T+1}=\sqrt{\mu_1}+\sqrt{\alpha-\mu_1}+\sqrt{\mu_2}+\sqrt{\beta-\mu_2}\;.
\end{equation}
It is trivial to see that the maximum of $n_T$ occurs when
$\mu_1=\mu_4=\alpha/2$ and $\mu_2=\mu_3=\beta/2$. This maximum cannot
be achieved, however, because we must respect the ordering $\mu_1
\geq \mu_2 \geq \mu_3 \geq \mu_4$ that we assumed in our definition
of $\hat{n}_{\Phi}$.  We should always choose $\mu_2=\mu_3=\beta/2$,
but the best we can then do with $\mu_1$ and $\mu_4$ is to choose
$\mu_4=\beta/2$, $\mu_1=\alpha-\beta/2$ when $\alpha\geq\beta$ [upper
sign in Eq.~(\ref{phialpha})] or $\mu_1=\beta/2$,
$\mu_4=\alpha-\beta/2$ when $\beta\geq\alpha$ [lower sign in
Eq.~(\ref{phialpha})].  The requirement that $\mu_1\le\alpha$ implies
that the latter case can only be used when $\hat n_\Phi\ge\sqrt2$. In
both cases, the the maximum value of $n_T$ for fixed $\hat{n}_{\Phi}$
has the form
\begin{equation}
\label{eq:doublyB5}
n_T = \frac{1}{2}\left[\left(\sqrt{\alpha - \beta/2} + 3\sqrt{\beta/2}\right)^2 - 1\right].
\end{equation}
It turns out that the upper sign in Eq.~(\ref{phialpha}) always gives
a larger value for $n_T$.  Using the upper sign, we find that the
maximum of $n_T$ for fixed values of $\hat{n}_{\Phi}$ is given by
\begin{equation}
\label{eq:doublyB5b}
n_T = \frac{3}{4}\!\left( 1- \sqrt{1 - \frac{4}{9} \hat{n}_{\Phi}^2} +
\sqrt{\frac{4}{3} \hat{n}_{\Phi}^2 + 2 \sqrt{1-\frac{4}{9} \hat{n}_{\Phi}^2} -2}
\; \right).
\end{equation}

\begin{figure}[!ht]
\resizebox{8 cm}{8 cm}{\includegraphics{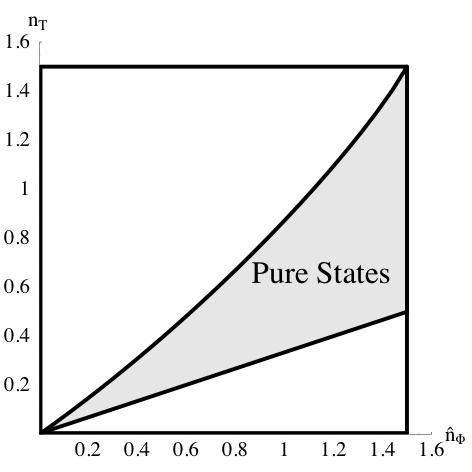}}
 \caption{The pure-state region in the $\hat{n}_{\Phi}$-$n_T$ plane for $4
\times N$
 systems.}
 \label{fig:pure}
\end{figure}

The minimum value of $n_T$ occurs on the boundary of allowed Schmidt
coefficients, i.e., when $\mu_1=\alpha$ and $\mu_2=\beta$, with the
upper sign in Eq.~(\ref{phialpha}).  Thus the minimum value of $n_T$
for a fixed value of $\hat{n}_{\Phi}$ is given by
\begin{equation}
\label{eq:doublyB8}
n_T =
\frac{1}{2}\left[(\sqrt{\alpha} + \sqrt{\beta}\,)^2 - 1\right]=
\frac{1}{3} \hat{n}_{\Phi}.
\end{equation}
From Eqs.~(\ref{eq:doublyB5}) and (\ref{eq:doublyB8}) we find that the
pure states of a $4\times N$ system lie in the region shown in Fig
\ref{fig:pure}.  Notice that for this case of two constraints, the
pure-state region is not convex.

\subsection{Entanglement of formation} \label{doubly:eof}

The EOF for pure bipartite states is a concave function of the marginal
density operator obtained by tracing over one of the subsystems. This
means that it is a concave function of the Schmidt coefficients
$\vec{\mu}$. Searching for a minimum is not the most natural thing one can
do with a concave function, yet this is what we are instructed to do by
the procedure for bounding the EOF outlined in Sec.~\ref{subsec:multiply}.
Starting from the EOF $H(\vec{\mu})$ for pure bipartite $4 \times N$
states, our objective is to find a convex, monotonic function ${\mathcal
H}({\bf n})$ as outlined in the Sec.~\ref{general}. This function will be
our lower bound on the EOF for all states.

The first step is to find the function
\begin{equation}
\label{eq:doublyC1}
\widetilde{H}({\bf n}) = \widetilde{H}\big(\hat{n}_{\Phi}, n_T \big) \equiv
\min_{\vec{\mu}} \left\{ H(\vec{\mu}) \left| 3 \sqrt{(\mu_1+
\mu_4)(\mu_2+\mu_3)} = \hat{n}_{\Phi}, \; \frac{ \left(\sum_j \sqrt{\mu_j}
\right)^2
-1 }{2} = n_T  \right. \right\},
\end{equation}
which is defined on the pure-state region.  The method of Lagrange multipliers
is not suitable for finding the minimum in Eq.~(\ref{eq:doublyC1}) because the
problem is over-constrained. The equations that we obtain using Lagrange
multipliers have a consistent solution only if $\hat{n}_{\Phi}$ and $n_T$ are
related as in Eq.~(\ref{eq:doublyB5b}) and therefore lie on the upper boundary
of the pure-state region. This does not mean that there is no minimum for
$H(\vec{\mu})$, but rather that the minimum lies on a boundary of the allowed
values of $\vec{\mu}$.

We already know $\widetilde{H}\big(\hat{n}_{\Phi},\,n_T \big)$ on
the boundaries of the pure-state region. The boundary with three
of the Schmidt coefficients being zero is the origin in the
$\hat{n}_{\Phi}$-$n_T$ plane where
$\widetilde{H}\big(\hat{n}_{\Phi},\,n_T \big)=H(\vec{\mu})=0$. The
boundary with two of the Schmidt coefficients zero lies on the
line $n_T=\hat{n}_{\Phi}/3$. To find the value of
$\widetilde{H}\big(\hat{n}_{\Phi},\, n_T \big)$ along this
boundary, note that the minimum of the EOF subject to just the
$\hat{n}_{\Phi}$ constraint occurs for $\vec{\mu}_{\Phi}=(\alpha,
1-\alpha,0,0)$, where $\alpha$ is given in Eq.~(\ref{phialpha}).
Substituting $\vec{\mu}_{\Phi}$ into $n_T$ we get $n_T=
\sqrt{\alpha (1-\alpha)}=\hat{n}_{\Phi}/3$. This means that along
the line $ n_T = \hat{n}_{\Phi}/3$, the $n_T$ constraint is
automatically satisfied if the $\hat{n}_{\Phi}$ constraint is
satisfied. Thus along the lower boundary of the pure-state region,
we have $\widetilde{H}\big(\hat{n}_{\Phi},\, n_T \big) =
\widetilde{H}\big( \hat{n}_{\Phi} \big)$. Similarly, along the
upper boundary of the pure-state region, the $\hat{n}_{\Phi}$
constraint comes for free. This is because the minimum of the EOF
subject to the $n_T$ constraint occurs when the Schmidt
coefficients are given by $\vec{\mu}_T = (\gamma, \gamma',
\gamma', \gamma')$ with $\gamma$ given by Eq.~(\ref{eq:gamma}) and
$\gamma' = (1-\gamma)/3$. The doubly-constrained problem reduces
to the singly-constrained  problem when
$\hat{n}_{\Phi}=\sqrt{2(2\gamma+1) (1-\gamma)}$. Relabelling
$\gamma$ as $\alpha - \beta/2$ and $\gamma'$ as $\beta/2$ we see
that the $\hat{n}_{\Phi}$ constraint is automatically satisfied
along the upper boundary of the pure-state region if the $n_T$
constraint is satisfied. Hence along the upper boundary of the
pure-state region, we have $\widetilde{H} \big(\hat{n}_{\Phi},\,
n_T \big)=\widetilde{H}\big( n_T \big)$.

These considerations mean that for the entanglement of formation, the
monotone boundaries that we define in Appendix~\ref{A:construction}
coincide with the boundaries of the pure-state region, making it
unnecessary to construct the monotonically nondecreasing function
$\widetilde H_{\uparrow}\big(\hat n_\Phi,n_T\big)$, since $\widetilde
H\big(\hat n_\Phi,n_T\big)$ is itself monotonically nondecreasing.

\begin{figure}[!ht]
\resizebox{8 cm}{8 cm}{\includegraphics{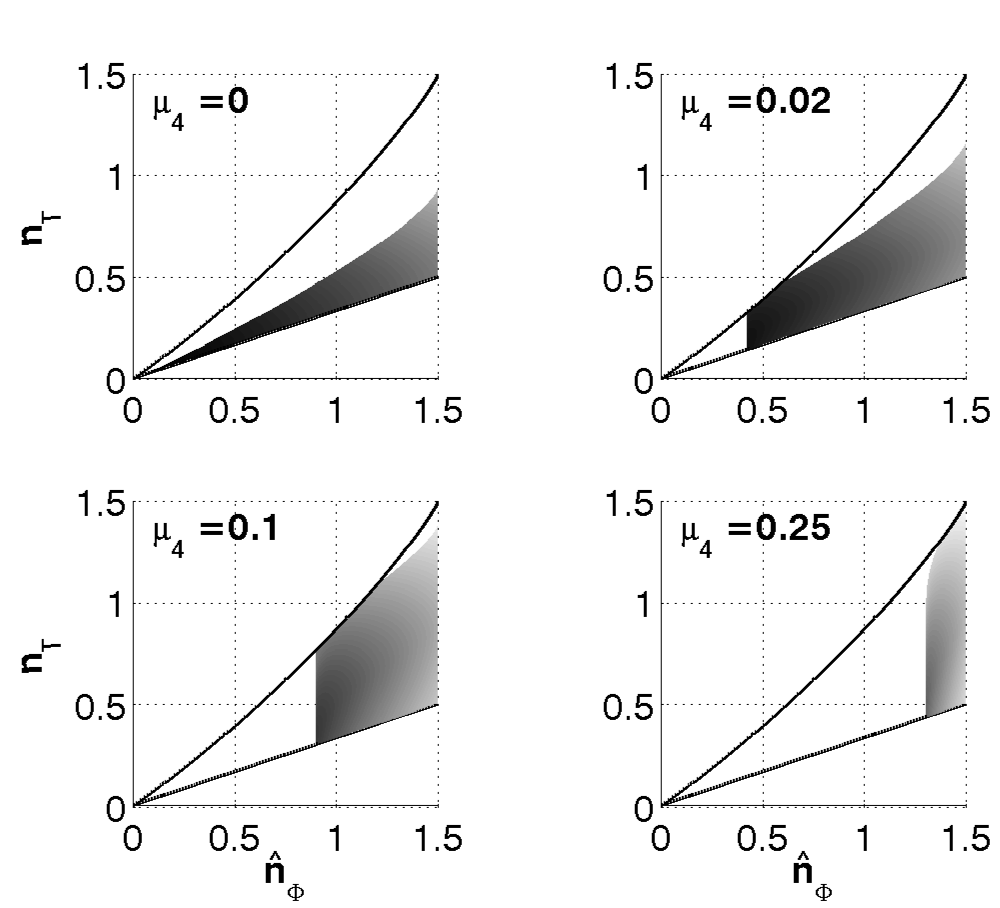}}
 \caption{The part of the 2-constraint region in which a value for
$\widetilde{H}({\bf n})$ can be computed is shown for four values of
$\mu_4=0$, 0.02, 0.1, and 0.25.  The two lines are the boundaries of
the pure-state region.}
 \label{fig:m4regions}
\end{figure}

The minimum of $H(\vec{\mu})$ in the remaining part of the pure-state
region can be found using the straightforward numerical procedure
described below. We start from the two distinct sets of solutions
$\vec{\mu}^{(1)}$ and $\vec{\mu}^{(2)}$ of the three constraint
equations (see Appendix~\ref{appB}). We go to the boundary where one
of the Schmidt coefficients is zero by setting $\mu_4=0$ in the
solutions. Now compute $H\big(\vec{\mu}^{(1)} \big)$ and
$H\big(\vec{\mu}^{(2)} \big)$ corresponding to the two solutions in
the regions in the $\hat{n}_{\Phi}$-$n_T$ plane where each of the
solutions is valid. The solutions are not valid in the whole
pure-state region because the three Schmidt coefficients have to be
real, nonnegative numbers less than one. All points in the pure-state
region cannot be covered if we set $\mu_4=0$. This is easily seen by
noticing that the point $\hat{n}_{\Phi}=n_T=3/2$ corresponds to the
fully entangled $4 \times N$ state and for this state all four
Schmidt coefficients have the value $1/4$. The fully entangled state
and other states close to it cannot be reached using the procedure
described above if we stay on the boundary defined by $\mu_4=0$. So
we start increasing the value of $\mu_4$ in small steps until it
reaches $1/4$. The parts of the 2-constraint region that are covered
by different choices of $\mu_4$ are shown in
Fig.~\ref{fig:m4regions}.

This numerical procedure gives us ranges of values of $\mu_4$ over which
$H\big(\vec{\mu}^{(1)} \big)$ and/or $H\big(\vec{\mu}^{(2)} \big)$ can be
calculated at each point in the pure-state region.  For the value of
$\widetilde{H}({\bf n})$ at each point, we pick the minimum over the
allowed range of values for $\mu_4$ at that point.

\begin{figure}[!ht]
\resizebox{16 cm}{8 cm}{\includegraphics{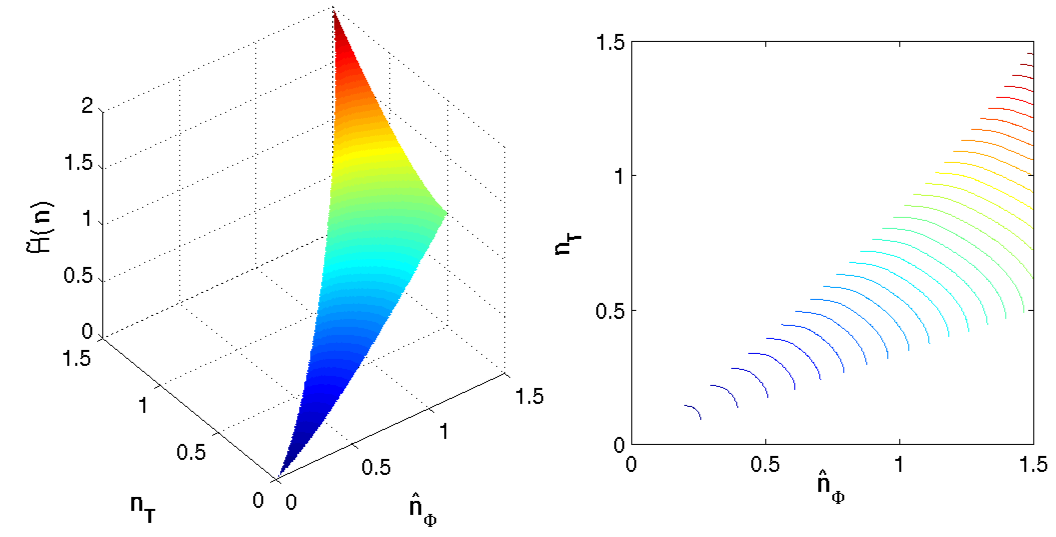}}
 \caption{(Color online) Plots of $\widetilde{H}({\bf n})$, the minimum of
the entropy of formation, $H(\vec{\mu})$, in the pure-state region. On
the left side is a 3-dimensional plot of $\widetilde{H}({\bf n})$ and on
the right is a contour plot of the same function.}
 \label{fig:eof2constraint}
\end{figure}

The function $\widetilde{H}({\bf n})$ in the pure-state region is
shown in Fig.~\ref{fig:eof2constraint}. It is, as required, a
monotonically increasing function of both $\hat{n}_{\Phi}$ and $n_{T}$. Along
the upper boundary of the pure-state region, the numerically
computed value of $\widetilde{H}({\bf n})$ matches the value of
$\widetilde{H}\big(n_T\big)$ from Eq.~(\ref{negmin}). In addition to this,
from the contour plot of $\widetilde{H}({\bf n})$ in
Fig.~\ref{fig:eof2constraint}, we see that along the upper  boundary, the
function has zero slope along the $\hat{n}_{\Phi}$ direction.

The function  $\widetilde{H}({\bf n})$ is not convex, which
can be seen by computing the Hessian at every point in the pure-state
region. If the function were convex, both eigenvalues of the Hessian would
be positive at all points. It turns out that one of the eigenvalues of the
Hessian is negative in a region in the upper right corner of the
$\hat{n}_{\Phi}$-$n_T$ plane, close to the maximally entangled state.

Since $\widetilde{H}({\bf n})$ is not convex, we have to compute its
convex hull,
\begin{equation}
\label{eq:doublyC7}
{\mathcal H}({\bf n}) = {\mbox{co}}\left[ \widetilde{H}({\bf n})
\right],
\end{equation}
to obtain the bound on the EOF in the pure-state region. The convex hull
of $\widetilde{H}({\bf n})$ can be computed numerically, and it
turns out that the difference between ${\mathcal H}({\bf n})$ and
$\widetilde{H}({\bf n})$ is quite small ($\sim 10^{-3}$), the
two differing differ only in a small region in the upper right corner of
the pure-state region.  As shown in Appendix~\ref{AppProof}, taking the
convex hull preserves monotonicity.

 \begin{figure}[!ht]
\resizebox{16 cm}{8 cm}{\includegraphics{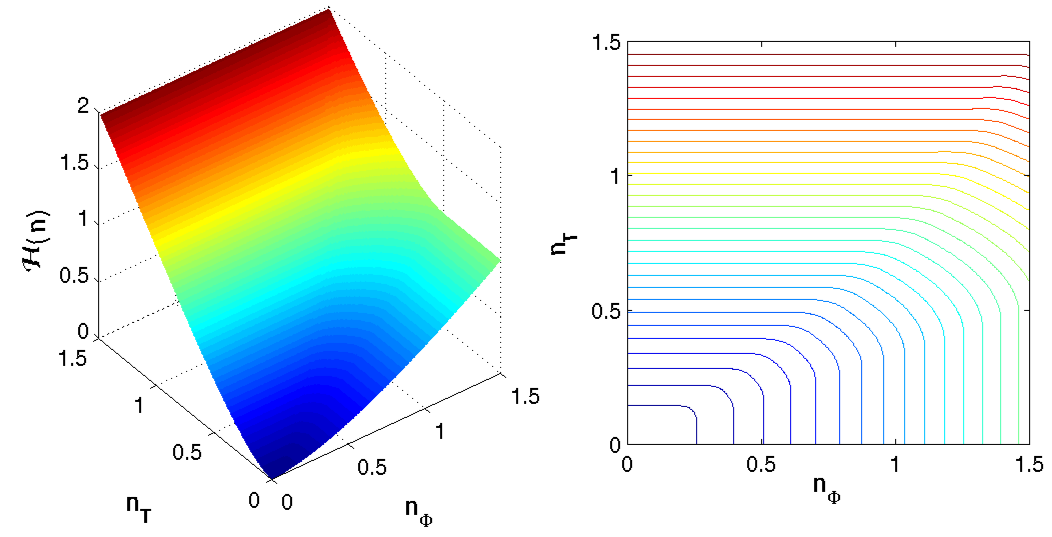}}
 \caption{(Color online) The doubly-constrained bound ${\mathcal H}({\bf n})$ on
the EOF of all $4 \times N$ states. On the right side is a contour plot of the
same function.}
 \label{fig:eofbound}
\end{figure}

To obtain a bound on the EOF of {\em all}\/ $4 \times N$ states, we have
to extend ${\mathcal H}({\bf n})$ out of the pure-state region to the rest
of the $\hat{n}_{\Phi}$-$n_T$ plane. The extension has to respect the
monotonicity of ${\mathcal H}({\bf n})$ so that the string of inequalities
Eq.~(\ref{eq:doublyA6}) holds. This can be achieved by extending
${\mathcal H}({\bf n})$ using surfaces that match the function at the
lower and upper boundaries of the pure-state region. To preserve
monotonicity, the surface added on to the region below the lower boundary
of the set of pure states has zero slope along the $n_T$ direction, and the
surface added on to the region above the upper boundary of the set of pure
states has zero slope along the $\hat{n}_{\Phi}$ direction.  The resulting
doubly-constrained bound  ${\mathcal H}({\bf n})$ on the EOF is shown in
Fig.~\ref{fig:eofbound}. We see from the figure that the extension to the
whole $n_{\Phi}$-$n_T$ plane produces a smooth and seamless surface.

One final point worth mentioning involves the use of our bound for general
mixed states.  To do so, one must calculate $n_\Phi$ for the mixed state, and
this calculation depends on the choice of an angular-momentum basis for system
$B$ in order to define the $\Phi$-map.  The bound itself thus depends on this
choice of basis, and the best bound would generally be found for the basis
choice that gives the largest value of $n_\Phi$.  For pure states, for example,
the results in Fig. \ref{bound} show that the best choice of basis is the
Schmidt basis for system $B$.

\begin{figure}[!ht]
\resizebox{7 cm}{7
cm}{\includegraphics{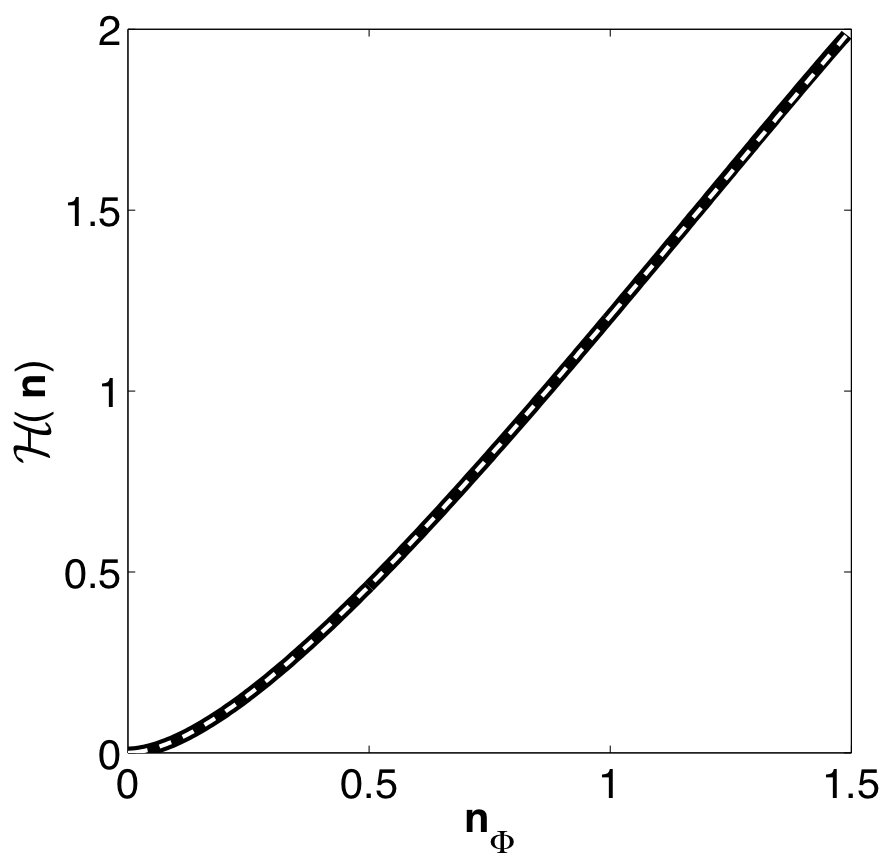}} \caption{The
thick black line is the doubly-constrained bound on the EOF for
isotropic states. The dashed white line, lying on top of the black
line, is the singly-constrained bound $\mathcal{H}\big(n_T\big)$
from Eq.~(\ref{Tbound}).}
 \label{fig:eofcompare}
\end{figure}

The isotropic states, which lie along the diagonal in the
$\hat{n}_{\Phi}$-$n_T$ plane, are special in that they saturate the
singly-constrained bound ${\mathcal H}\big(n_T \big)$ from
Eq.~(\ref{Tbound}). These states thus furnish a good consistency test
of our doubly-constrained bound because our bound must match the
singly-constrained bound when applied to isotropic states. A
comparison of the two bounds for isotropic states is given in
Fig.~\ref{fig:eofcompare}.

We can make a second comparison between the singly and doubly
constrained bounds using Fig.~\ref{fig:eofcompare}. From the way we
constructed ${\mathcal H}({\bf n})$, we know that its value on the
diagonal in the $\hat{n}_{\Phi}$-$n_T$ plane is the same as its value
on the upper boundary of the pure-state region. We also know that the
upper boundary is where the singly-constrained bound and the
doubly-constrained bound are the same. From
Fig.~\ref{fig:eofcompare}, we see that the convex hull
$\mathcal{H}\big(n_T\big)$ of the function $\widetilde{H}\big( n_T
\big)$ of one variable matches the convex hull ${\mathcal H}(\bf n)$
of the function $\widetilde{H}({\bf n})$ of two variables on the
upper pure-state boundary.  These consistency checks give us
increased confidence in the accuracy of our results.

\subsection{Tangle and concurrence}\label{doubly:tangle}

Doubly-constrained bounds can be placed on the tangle and the
concurrence of $4 \times N$ states by extending the procedure used
for the EOF. For the tangle, we start by finding the function,
\begin{equation}
\label{eq:doublyD1}
\widetilde{T}({\bf n}) = \min_{\vec{\mu}} \left\{ 2\left( 1- \left| \vec{\mu}
\right|^2 \right) \left| 3\sqrt{(\mu_1 + \mu_4)(\mu_2 + \mu_3)} =
\hat{n}_{\Phi},
\frac{\left( \sum_j \sqrt{\mu_j} \right)^2-1}{2} = n_T \right. \right\},
\end{equation}
in the pure-state region.  For the concurrence, we want the function
$\widetilde C({\bf n})= \sqrt{\widetilde T({\bf n})}$, since for pure
states the concurrence is the square root of the tangle.

 \begin{figure}[!ht]
\resizebox{16 cm}{8 cm}{\includegraphics{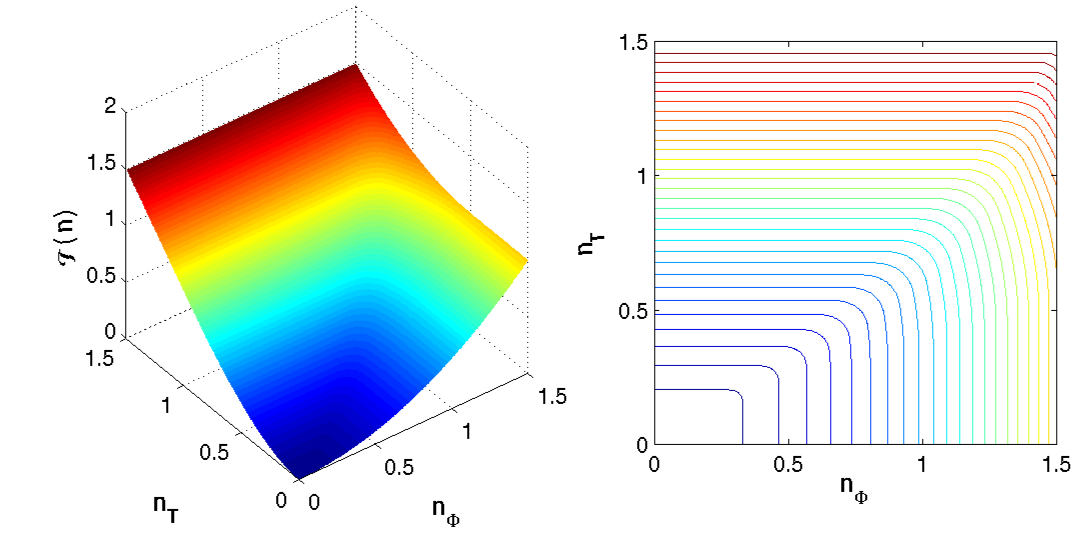}}
 \caption{(Color online) The doubly-constrained bound ${\mathcal T}({\bf n})$ on
the tangle of  $4 \times N$ states. On the right side is a contour plot of the
same function.}
 \label{fig:tanglebound}
\end{figure}

\begin{figure}[!htb]
\resizebox{16 cm}{8 cm}{\includegraphics{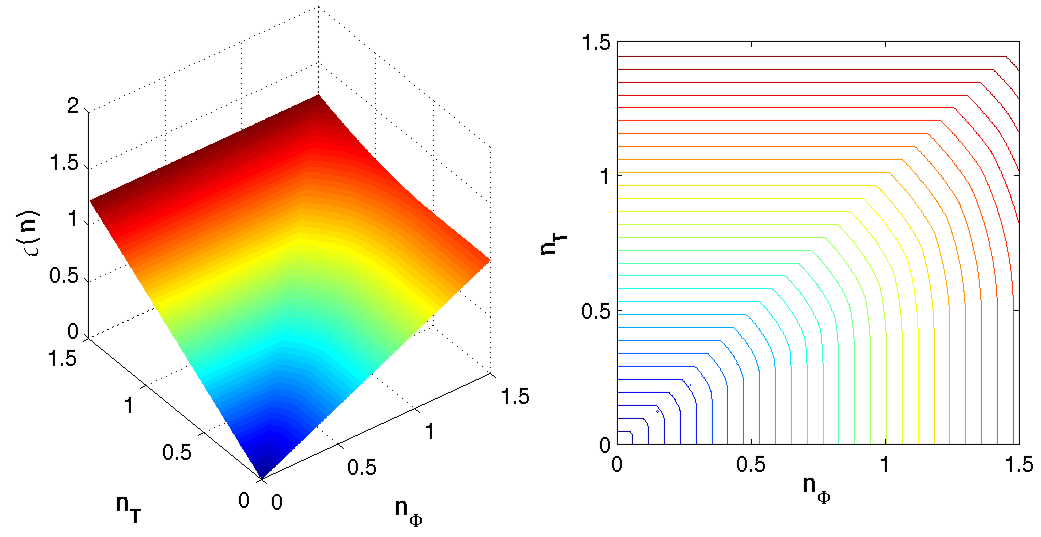}}
 \caption{(Color online) The doubly-constrained bound ${\mathcal C}({\bf n})$ on
the concurrence of $4 \times N$ states. On the right side is a
contour plot of the same function.}
 \label{fig:concbound}
\end{figure}

For all three of the entanglement monotones, EOF, tangle and
concurrence, the monotone boundaries we define in
Appendix~\ref{A:construction} coincide with the boundaries of the
pure-state region.  This is because the singly-constrained bounds for
all three measures correspond to the same sets of Schmidt
coefficients, $\vec{\mu}_T=(\gamma,\gamma',\gamma', \gamma')$ and
$\vec{\mu}_{\Phi}=(\alpha, 1-\alpha, 0, 0)$, and we have already seen
for the EOF that these Schmidt coefficients define the boundaries of
the pure-state region.   This makes it unnecessary for these
entanglement monotones to construct the monotonically nondecreasing
function discussed in Appendix~\ref{A:construction}.  In general, for
two different measures of entanglement and two constraints, the
singly-constrained bounds for the two measures need not correspond to
the same Schmidt coefficients.

Once we have $\widetilde{T}({\bf n})$, the convex hull of this function
extended to the whole $\hat{n}_{\Phi}$-$n_T$ plane is the doubly-constrained
bound on the tangle, ${\mathcal T}({\bf n})$. A three-dimensional plot and a
contour plot of ${\mathcal T}({\bf n})$ are shown in Fig.~\ref{fig:tanglebound}.

The bound on the concurrence is the convex hull of the surface
obtained from $\widetilde C({\bf n})$. The resulting bound on the
concurrence, ${\mathcal C}({\bf n})$, is shown in
Fig.~\ref{fig:concbound}.

\section{Conclusion}\label{conclusion}

We focused on two aspects of the problem of quantifying entanglement in
this paper. The first was a comparison between the bounds on different
measures of entanglement obtained by using $n_T$ and $\hat{n}_{\Phi}$
independently as constraints. The second was the construction of
doubly-constrained bounds on the three measures of entanglement that we
considered.

Starting from the $\Phi$-map~\cite{Breuer2006}, we found that we
can define an entanglement measure, which we call the
$\Phi$-negativity. The $\Phi$-negativity of arbitrary quantum
states can be calculated in a straightforward manner, just like
their negativity. We also found that we can obtain a much simpler
function $\hat{n}_{\Phi}$ of the Schmidt coefficients of pure
states that is an upper bound on their $\Phi$-negativity. Previous
work~\cite{Chen2005a},\cite{Chen2005} has shown that the
negativity can be used as a constraint to place bounds on the EOF,
the tangle, and the concurrence of bipartite states. We obtained a
different set of bounds on these three measures of entanglement
for $4 \times N$ mixed states by using $\hat{n}_{\Phi}$ instead as
the constraint. The scheme for placing lower bounds on
nonoperational measures of entanglement is general enough to allow
us to use $\hat{n}_{\Phi}$ instead of $n_{\Phi}$ as the
constraint. We were then able to compare the two sets of bounds on
the measures of entanglement coming from using either one of the
two operational entanglement measures as a single constraint.

We found that the $\hat{n}_{\Phi}$-$n_T$ plane for pure states can
be divided into two regions depending on which constraint led to
the better bound on a given measure of entanglement. This prompted
us to consider whether we can construct a single, composite bound
for each measure of entanglement, applicable to the entire
$\hat{n}_{\Phi}$-$n_T$ plane, by using both constraints
simultaneously. It turned out that for $4 \times N$ systems this
is a tractable problem, and we obtained doubly-constrained lower
bounds for the first time for the EOF, the tangle, and the
concurrence. We showed how the bounds on the different measures of
entanglement obtained for pure states can be extended to include
all states. We found that the requirement of monotonicity on the
bound defined on pure states dictates how to extend the bound to
all states.

\appendix

\section{General Construction of \texorpdfstring{$\widetilde{G}_{\uparrow}
({\bf n})$}{G (up)}.}\label{A:construction}

In this Appendix, we describe the general procedure for constructing the
monotonically nondecreasing function $\widetilde{G}_{\uparrow} ({\bf n})$,
which replaces $\widetilde{G}({\bf n})$ when the latter function is not
itself monotonically nondecreasing.

As mentioned in Sec \ref{subsec:multiply}, pure states of the
system correspond to a simply connected subset, called the
pure-state region, in the state hypercube in $\mathbb{R}^K$; the
function $\widetilde{G}({\bf n})$ is defined only on the
pure-state region.  Within the pure-state region, we can define
$K$ hypersurfaces ${\mathcal S}_k$ as those on which the $k$th
constraint equation, $F_k(\vec{\mu})=n_k$, is automatically
satisfied if the remaining $K-1$ constraint equations are
satisfied.  We denote the value of $n_k$ on ${\mathcal S}_k$ by
$n^*_k({\bf n'})$ where ${\bf n'} = (n_1,\ldots ,n_{k-1}, n_{k+1},
\ldots,n_K)$; the function $n^*_k({\bf n'})$ can be regarded as
the defining equation for $S_k$.  On the hypersurfaces $S_k$,
$\widetilde{G}({\bf n})$ is effectively defined by $K-1$
constraints. We denote the value of $\widetilde{G}({\bf n})$ on
${\mathcal S}_k$ by $\widetilde{G}_k({\bf n'})$. The minimum of
any function subject to $K$ constraints is always greater than or
equal to its value when subject to $K-1$ of these constraints, so
we have $\widetilde{G}({\bf n'}, n_k ) \geq \widetilde{G}_k({\bf
n'})$, where we have let $({\bf n'}, n_k) \equiv {\bf n}$. The
inequality is saturated when $n_k = n_k^*({\bf n'})$. Now consider
$\widetilde{G}({\bf n'}, n_k )$ as a function of $n_k$. If we fix
${\bf n'}$ and increase $n_k$, starting from its lowest value,
then $\widetilde{G}({\bf n'}, n_k )$ has to either decrease or
remain constant until we cross the hypersurface ${\mathcal S}_k$.
For $n_k \geq n_k^*({\bf n'})$, $\widetilde{G}({\bf n'}, n_k )$ is
a nondecreasing function of $n_k$. We want
$\widetilde{G}_{\uparrow} ({\bf n})$ to be a nondecreasing
function for all $n_k$, so we define it by
\begin{equation}
\label{eq:g_up1} \widetilde{G}_{\uparrow}({\bf n'}, n_k) = \bigg\{
\begin{array}{cl}
\widetilde{G}_k({\bf n'}) &  n_k \leq n_k^*({\bf n'}) \\
\widetilde{G}({\bf n}) &  n_k > n_k^*({\bf n'}),\end{array} \quad
k=1,\ldots, K.
\end{equation}
The construction of $\widetilde{G}_{\uparrow} ({\bf n})$ is not
complete at this point. Within each ($K-1$)-dimensional
hypersurface, we will encounter ($K-2$)-dimensional hypersurfaces
where two of the constraints are automatically satisfied.  Across
each of these ($K-2$)-dimensional hypersurfaces, we can update the
value of $\widetilde{G}_{\uparrow} ({\bf n})$ just as described
above.

There can be at most $K$ different $(K-1)$-constraint regions and the
$K$-constraint region will, in general, be surrounded by these
$(K-1)$-constraint regions. The $(K-1)$-constraint regions are
surrounded, in turn, by $(K-2)$-constraint regions and so on. This
construction procedure evidently terminates after $K-1$ steps.  We
call the hypersurfaces identified in this appendix~\textit{monotone
boundaries}, because the nested structure of $k$-constraint regions
they define are the key to constructing the monotonically
nondecreasing function $\widetilde{G}_{\uparrow} ({\bf n})$ from
$\widetilde{G}({\bf n})$.  In the examples we consider in
Sec.~\ref{doubly}, the monotone boundaries coincide with the
boundaries of the pure-state region, so we do not have to construct
the function $\widetilde G_{\uparrow}({\bf n})$.

\section{The convex hull and monotonicity}\label{AppProof}

Here we show that the convex hull of a monotonically nondecreasing
function on $\mathbb R^K$ is also monotonically nondecreasing.  We
first define a partial order on the set of vectors in $\mathbb
R^K$ by defining ${\bf x} \ge {\bf y}$ to mean $x_k \ge y_k$ for
all $k$.  Define a {\it monotone\/} to be a function
$f:\mathcal{D}\mapsto [0,1]$ satisfying the following conditions:
\begin{enumerate}
\item The domain $\mathcal{D}$ is a bounded region contained in
the positive orthant (including boundaries) of $\mathbb{R}^K$,
\item $0 \in \mathcal{D}$ \mbox{and} $f(0)=0$, \item $\forall \;\;
{\bf x},{\bf y} \in \mathcal{D}$, if ${\bf x} \ge {\bf y}$, then
$f({\bf x}) \ge f({\bf y})$ . \quad \mbox{(monotonicity)}
\end{enumerate}
The function $f$ can alternatively be viewed as a set of points in
$\mathbb R^{K+1}$ given by the tuples $(x_1, \ldots, x_K, f({\bf
x}))$. Viewed this way, we can define the convex hull $C$ of $f$
as a \emph{set\/} to be the smallest convex set containing the set
$f$.  We can also define the function $c:\mathcal D^\prime \mapsto
[0,1]$, to be the convex hull of $f$ as a \emph{function\/}.  Thus
$c$ is the largest convex function bounded from above by $f$; in
this paper $c$ is called the convex roof of $f$;   Clearly, $c$ is
just the lower boundary of the set $C$ along the direction of the
$(K+1)$st coordinate in $\mathbb R^{K+1}$.

Before continuing to the main theorem, we state an important
result known as {\it Carath\'eodory's theorem\/}
\cite{Rockafellar1997}.  This theorem uses the notion of a {\it
generalized simplex of dimension $d$\/}, which is just the convex
hull of a set of $d+1$ affinely independent points.  A triangle,
or example, regardless of shape, is a generalized simplex of
dimension 2. For convenience we refer to a generalized simplex as
just a simplex.

\begin{theorem}[Carath\'eodory]
Let $f$ be any bounded set of points in $\mathbb R^{K+1}$, and let
$C = {\rm co}[f]$ be the convex hull of $f$ (as a set).  Then
${\bf x} \in C$ if and only if ${\bf x}$ can be written as a
convex combination of $K+2$ (not necessarily distinct) points in
$f$.  Furthermore, $C$ is the union of all the simplices with
dimension less than or equal to $K+1$ whose vertices belong to
$f$.
\end{theorem}

From Carath\'eodory's theorem and the fact that the function $c$
is the boundary of the set $C$, we know that $c$ can be expressed
as the union of many simplices (usually infinitely many) whose
vertices belong to $f$. These simplices are necessarily of
dimension at most $K$, since the dimension of $c$ is $K$.  We can
speak meaningfully about directional derivatives on these
simplices and on $c$ because of the following beautiful fact: any
convex function has well defined one-sided directional derivatives
everywhere and, furthermore, is differentiable everywhere except
possibly a set of measure zero~\cite{Rockafellar1997}.

\begin{theorem} Let $f:\mathcal D \mapsto [0,1]$ be a monotone, and let
$c:\mathcal D^\prime \mapsto [0,1]$ be the convex roof of the
function $f$. Then $c$ is also a monotone.  \end{theorem} {\it
Proof:\/} The domain $\mathcal D^\prime$ of $c$ in general
contains the domain $\mathcal D$ of $f$, but it will remain
bounded and in the positive orthant of $\mathbb R^{K}$ and is
furthermore always convex even if $\mathcal D$ is not.  Clearly $0
\in \mathcal D^\prime$, since $0 \in \mathcal D$.  The fact that
$c(0)=0$ can be seen by the fact that $f(0)=0$ is the global
minimum for $f$, and the convex hull of a function will always
contain the function's global minimum.  This shows that $c$
satisfies the first two criteria of a monotone.

Now we prove the final criterion, the monotonicity of $c$.
Consider the set of all possible simplices with dimension less
than or equal to $K$ with vertices lying in $f$.  From
Carath\'eodory's theorem, $c$ is a union of some subset of these
simplices. However, every simplex in this set has the property of
monotonicity over its domain of definition.  This follows from the
``multidirectional'' version of the mean value theorem
\cite{Clarke1994},\cite{Clarke1994a}, for which we now sketch the
proof. Suppose we choose a simplex $s \in c$.  Along a given
direction ${\bf p}$, the smallest value of the directional
derivative of $f$ lying above $s$ is a lower bound on the
directional derivative of $s$.  In particular, if ${\bf p} \ge 0$,
then from the assumption of monotonicity of $f$, we know that
$\nabla_{{\bf p}} f \ge 0$ everywhere, and hence  $\nabla_{{\bf
p}} s \ge 0$.  This implies that each constituent simplex in $c$
is indeed monotonic.  To show that $c$ is a monotone, we use the
convexity of $c$ to see that the directional derivative in some
direction ${\bf p} \ge 0$ across two neighboring simplices $s_1$
and $s_2$ cannot decrease.

\section{\texorpdfstring{$\hat{n}_{\Phi}$}{n-phi} for
\texorpdfstring{$D\times N$}{D by N} Pure States} \label{appA}

Our objective in this Appendix is to characterize the eigenvalues of
the operator $(I\otimes \Phi)\rho_{AB}=\mathcal{O}$ in
Eq.~(\ref{phionrho}) for the special case in which the Schmidt basis
for subsystem $B$ of the pure state $\rho_{AB}$ is the same as the
angular-momentum basis.  Recall that in Eq.~(\ref{phionrho}), the
Schmidt coefficients $\mu_i$ are ordered from largest to smallest.
The operator $\mathcal{O}$ is a $DN \times DN$ operator, although it
clearly has rank at most $D^2$, so we can regard it as a $D^2\times
D^2$ operator, having $D^2$ eigenvalues. Although $\mathcal{O}$ can
be written in matrix form in the Schmidt basis, we refrain from doing
so here, as the expression is unwieldy and not very illuminating.  We
can, however, by permuting the rows and columns of $\mathcal{O}$,
write it as
 \begin{equation}
\mathcal{O} = \mathbf{0} \oplus \mathbf{T} \oplus \mathbf{R},
 \end{equation}
where $\mathbf{0}$ is a matrix of zeros, of size $D \times D$.

To describe $\mathbf T$ and $\mathbf R$, we first make some
definitions. An {\it index\/} is an integer between 1 and $D$. An
ordered pair of indices $(j,k)$ is said to be {\it inadmissible\/}
if $k=D-j+1$ or $k=j$. All other indices are said to be {\it
admissible\/}.  A set of indices is called admissible if the
elements are pairwise admissible.  A product of $n$ distinct
Schmidt coefficients $\mu_{j_1} \mu_{j_2} \cdots \mu_{j_n}$ is
said to be {\it $n$-admissible\/} if all of the indices are
pairwise admissible, and if, in addition, $j_1 < j_2 < \ldots <
j_n$. Finally, define $\mathcal{S}_n$ as the sum over all
$n$-admissible products. Then
 \begin{equation}
 \mathbf{T} = \hspace{-3 mm} \bigoplus_{\stackrel{(p,q)} {\mathrm{admissible}}}
\hspace{-3 mm} W_{(p,q)},
 \end{equation}
where each $W_{(p,q)}$ is a $2 \times 2$ matrix of the form
 \begin{equation}
 W_{(p,q)} = \left(%
\begin{array}{cc}
  \mu_p & (-1)^{p+q-1}\sqrt{\mu_p\mu_q} \\
  (-1)^{p+q-1}\sqrt{\mu_p\mu_q} & \mu_q \\
\end{array}%
\right).
 \end{equation}
For each index, there are $D-2$ other indices with which it can
form an admissible pair. Hence, $D$ indices form exactly
$D(D-2)/2$ distinct admissible pairs, and that is the number of
possible $W_{(p,q)}$'s of the given form. $W_{(p,q)}$ has
eigenvalues $0$ and $\mu_p + \mu_q$. Thus, $\mathbf{T}$ has
$D(D-2)/2$ zero eigenvalues and an equal number of eigenvalues
$\mu_p + \mu_q$, where $(p,q)$ is an admissible pair.

The matrix $\mathbf{R}$ has elements
 \begin{equation}
\mathbf{R}_{jk} = -\sqrt{\mu_j \mu_k}
(1-\delta_{j,k})(1-\delta_{j,D-k+1}),
 \end{equation}
where $j,k = 1,\cdots, D$. It is thus a $D \times D$ matrix. The
characteristic polynomial of this matrix can be written as
\begin{equation}
g(z)=z^{D/2}\left(z^{D/2}+ \sum_{t=0}^{D/2-2} t(-1)^t
\mathcal{S}_{t+1} z^{D/2-t-1} - (D/2-1) (-1)^{D/2} \mathcal{I}
\right) ,
\end{equation}
where
\begin{equation}
\mathcal{I}=\prod_{j=1}^{D/2} (\mu_j + \mu_{D-j+1}) .
\end{equation}
It is evident that the matrix $\mathbf{R}$ has $D/2$ zero
eigenvalues. The remaining eigenvalues are the zeroes of the
function
 \begin{equation}
 \label{rfunc}
r_D(z) = z^{D/2}+ \sum_{t=0}^{D/2-2} t(-1)^t \mathcal{S}_{t+1}
z^{D/2-t-1} - (D/2-1) (-1)^{D/2} \mathcal{I}.
 \end{equation}
The Descartes rule of signs tells us that the above equation has no
more than one negative root. In fact, if all the Schmidt coefficients
are nonzero, there is \emph{exactly one\/} negative eigenvalue, the
negative root of $r_D(z)$. Otherwise, all the eigenvalues are
nonnegative, and the pure state under consideration could be
separable.

Putting all this together, we conclude that the spectrum of
$\mathcal{O}$ has
 \begin{enumerate}
 \item $D + D(D-2)/2 + D/2 = D(D+1)/2$ zero eigenvalues,
 \item $D(D-2)/2$ positive eigenvalues of the form $\mu_p+\mu_q$, where $(p,q)$
is an admissible pair, and $D/2-1$ positive eigenvalues, which are
the positive roots of $r_D(z)=0$,
 \item One negative eigenvalue, the negative root of $r_D(z)=0$.
 \end{enumerate}

As an example, we present the case of $D=4$. Then Eq.~(\ref{rfunc})
becomes $r_4(z)\equiv z^2-(\mu_1+\mu_4)(\mu_2+\mu_3)$, which has
zeroes $\pm\sqrt{(\mu_1+\mu_4)(\mu_2+\mu_3)}$.

For $D=6$, the function~(\ref{rfunc}) is
 \begin{eqnarray}
 r_6(z)\equiv z^3 &-&z(\mu_1 \mu_2 + \mu_1 \mu_3 + \mu_2 \mu_3 + \mu_1 \mu_4 +
\mu_2 \mu_4 + \mu_1 \mu_5  + \mu_3 \mu_5 + \mu_4 \mu_5 \nonumber\\
&+& \mu_2 \mu_6 + \mu_3 \mu_6 + \mu_4 \mu_6 + \mu_5 \mu_6)  +
2(\mu_1 + \mu_6)(\mu_2 + \mu_5)(\mu_3 + \mu_4).
 \end{eqnarray}

\section{Solutions of the constraint equations} \label{appB}

The three constraint equations,
\begin{eqnarray}
\label{eq:appB1} \frac{1}{2} \left[ \left( \sqrt{\mu_1} +
\sqrt{\mu_2} + \sqrt{\mu_3} +
\sqrt{\mu_4} \right)^2 -1 \right] & = & n_T, \nonumber \\
3 \sqrt{(\mu_1 + \mu_4)(\mu_2 + \mu_3)} &=& \hat{n}_{\Phi}, \nonumber \\
\mu_1 + \mu_2 + \mu_3 + \mu_4 &=&1,
\end{eqnarray}
can be solved to express $\mu_1$, $\mu_2$, and $\mu_3$ in terms of
$n_T$, $\hat{n}_{\Phi}$, and $\mu_4$. There are four sets of
solutions, of which only two are distinct because the other two
can be obtained by exchanging $\mu_2$ and $\mu_3$. The constraint
equations are invariant under this exchange. The two distinct
solutions are the following:
\begin{eqnarray}
\label{eq:s1} \mu_1^{(1)} &=& \frac{1}{2} \left(1+\sqrt{1-
\frac{4}{9} \hat{n}_{\Phi}^2}-2
\mu_4\right), \nonumber \\
\mu_2^{(1)} &=& \frac{1}{4} \left(1 - \sqrt{1- \frac{4}{9} \hat{n}_{\Phi}^2} + 2
\sqrt{{\mathcal G}_0  - {\mathcal G}_1\!\left(\mu_1^{(1)} \right) } \right),
\nonumber \\
\mu_3^{(1)} &=& \frac{1}{4} \left(1 - \sqrt{1- \frac{4}{9} \hat{n}_{\Phi}^2} -
2 \sqrt{{\mathcal G}_0  - {\mathcal G}_1\!\left( \mu_1^{(1)}
\right) } \right),
\end{eqnarray}
and
\begin{eqnarray}
\label{eq:s2} \mu_1^{(2)} &=& \frac{1}{2} \left(1-\sqrt{1- \frac{4}{9}
\hat{n}_{\Phi }^2}-2 \mu_4\right), \nonumber \\
\mu_2^{(2)} &=& \frac{1}{4} \left(1 + \sqrt{1- \frac{4}{9} \hat{n}_{\Phi
}^2} + 2 \sqrt{{\mathcal G}_0  - {\mathcal G}_1\!\left(\mu_1^{(2)}
\right) } \right),
\nonumber \\
\mu_3^{(2)} &=& \frac{1}{4} \left(1 + \sqrt{1- \frac{4}{9} \hat{n}_{\Phi}^2} - 2
\sqrt{{\mathcal G}_0  - {\mathcal G}_1\!\left( \mu_1^{(2)}
\right) } \right).
\end{eqnarray}
Here ${\mathcal G}_0$ and ${\mathcal G}_1$ are given by
\begin{equation}
\label{eq;s1a} {\mathcal G}_0 =  1 + 8 (n_T + \mu_4) \sqrt{\mu_4(2
n_T+1)} - 4n_T(n_T+4 \mu _4)-3 \mu_4(\mu_4+2),
\end{equation}
\begin{eqnarray}
\label{eq:s1b}
{\mathcal G}_1 \big( \mu_1 \big) & = & \frac{\mu_1^{2}}{12} +
\bigg(\frac{2 \mu_1}{3}\bigg)^{3/2} \left(\sqrt{2n_T + 1}-\sqrt{\mu _4}\right)
+ \frac{\mu_1}{3}
\left(3+8 n_T-8 \sqrt{\mu_4(2 n_T + 1)} +5 \mu _4\right) \nonumber \\
&& + \frac{4 \sqrt{6 \mu_1}}{3}  \left[\sqrt{\mu_4} \left( 1 - 2
\sqrt{\mu_4(2n_T + 1)} + \mu_4\right)-n_T \left(\sqrt{2 n_T + 1}-3
\sqrt{\mu _4}\right)\right] \vphantom{ \frac{\mu _1^{3/2}}{4}}.
\end{eqnarray}

\section*{Acknowledgements}

The authors thank M.~Horodecki, A.~Denney, S.~Merkel, and A.~Silberfarb
for useful discussions.  This work was supported in part by Office of Naval
Research Contract No.~N00014-03-1-0426.

\bibliography{eof}

\end{document}